\documentclass[fleqn,usenatbib]{mnras}

\usepackage{newtxtext,newtxmath}

\usepackage[T1]{fontenc}
\usepackage{ae,aecompl}

\usepackage{graphicx}	
\usepackage{amsmath}	
\usepackage{amssymb}	

\newcommand{\scikit}{{\sc scikit-learn}}
\newcommand{\lya}{Ly$\alpha$}
\newcommand{\HI}{\ion{H}{I}}

\title[Detecting neutral hydrogen at $z\gtrsim 3$]{Detecting neutral hydrogen at $z\gtrsim 3$ in large spectroscopic surveys of quasars}

\author[Fumagalli et al.]{
Michele Fumagalli$^{1,2,3}$, Sotiria Fotopoulou$^{4,3}$, Laura Thomson$^{3}$\\
  $^{1}$Dipartimento di Fisica G. Occhialini, Universit\`a degli Studi di Milano Bicocca, Piazza della Scienza 3, 20126 Milano, Italy \\
  $^{2}$Institute for Computational Cosmology, Durham University, South Road, Durham, DH1 3LE, UK \\
  $^{3}$Centre for Extragalactic Astronomy, Durham University, South Road, Durham, DH1 3LE, UK \\
  $^{4}$ HH Wills Physics Laboratory, University of Bristol, Tyndall Avenue, Bristol BS8 1TL, UK\\
}

\pubyear{\the\year}

\begin{document}
\label{firstpage}
\pagerange{\pageref{firstpage}--\pageref{lastpage}}
\maketitle

\begin{abstract}
We present a pipeline based on a random forest classifier for the identification of high column-density clouds of  neutral hydrogen (i.e. the Lyman limit systems, LLSs) in absorption within large spectroscopic surveys of $z\gtrsim 3$ quasars. We test the performance of this method on mock quasar spectra that reproduce the expected data quality of the Dark Energy Spectroscopic Instrument (DESI) and the WHT Enhanced Area Velocity Explorer (WEAVE) surveys, finding $\gtrsim 90\%$ completeness and purity for $N_{\rm HI} \gtrsim 10^{17.2}~\rm cm^{-2}$ LLSs against quasars of $g < 23~$mag at $z\approx 3.5-3.7$. 
After training and applying our method on 10,000 quasar spectra at $z\approx 3.5-4.0$ from the Sloan Digital Sky Survey (Data Release 16), we identify $\approx 6600$ LLSs with $N_{\rm HI} \gtrsim 10^{17.5}~\rm cm^{-2}$ between $z\approx 3.1-4.0$ with a completeness and purity of $\gtrsim 90\%$ for the classification of LLSs. Using this sample, we measure a number of LLSs per unit redshift of  $\ell(z) = 2.32 \pm 0.08$ at $z=[3.3,3.6]$. We also present results on the performance of random forest for the measurement of the LLS redshifts and \HI\ column densities, and for the identification of broad absorption line quasars. 
\end{abstract}




\section{Introduction}

Gas around and between galaxies, within the circumgalactic and intergalactic medium (CGM and IGM), plays a relevant role in regulating the activity of galaxies  \citep[e.g.][]{tumlinson2017}, is a substantial reservoir of baryons at all redshifts \citep[e.g.][]{fukugita2004}, and retains memory of the thermal and chemical evolution of the Universe \citep[e.g.][]{schaye2000,simcoe2011}. 
In the past decades, dedicated studies started at 3m telescopes and continued after the deployment of high-resolution spectrographs at 8m class telescopes have exploited the analysis of absorption lines imprinted in quasar spectra to map the physical properties (temperature, density, metallicity) of the IGM and CGM \citep[e.g.][]{steidel1992,prochaska2002,rafelski2014,neeleman2015,fumagalli2016,dodorico2016}. 
Multiple efforts have also been dedicated to understanding the connection between the gas detected in absorption and the galaxies seen in emission \citep[e.g.][]{steidel2010,bielby2011,turner2014}, a field that is rapidly growing thanks to the deployment of wide integral field spectrographs at 8m telescopes \citep[e.g.][]{peroux2011,schroetter2016,fossati2019,lofthouse2020}.

Large spectroscopic surveys of quasars and in particular the Sloan Digital Sky Survey \citep[SDSS;][]{york2000} have played a critical role for the advancement of this field. With hundreds of thousands of spectroscopically confirmed quasars, it has been possible to compile large catalogues of intervening absorption line systems \citep[e.g.][]{noterdaeme2009,prochaska2010,zhu2013,garnett2017}, an effort which has spurred a plethora of follow-up studies at 8~m telescopes. Furthermore, despite the limited quality of the individual spectra, it has been possible to leverage the statistical power of SDSS to advance our appreciation of how gas and dust are distributed around galaxies \citep[e.g.][]{menard2010,lan2014}, to map the distribution and physical properties of the IGM with redshift \citep[e.g.][]{becker2013}, and also to use the IGM as a tool for cosmology \citep[e.g.][]{slosar2013}. 

The efforts of compiling catalogues in ever growing spectroscopic samples has led to an increase in sophistication of the algorithms used to identify features that are characteristic of a particular family of astrophysical objects, such as \ion{Mg}{II}, \ion{C}{IV} and \ion{Si}{II} absorbers, damped Ly$\alpha$ absorbers (DLAs), or broad absorption line (BAL) quasars. 
With quasar samples surpassing the 100,000 mark, artificial intelligence and in particular machine learning techniques are becoming an increasingly popular way to classify spectra \citep{garnett2017,parks2018,guo2019,ho2020}. The need for reliable and fast classification is also becoming a requirement and a necessity as surveys using the Dark Energy Spectroscopic Instrument \citep[DESI][]{desi2016} and the WHT Enhanced Area Velocity Explorer \citep[WEAVE][]{dalton2012} will grow the size of spectroscopic samples of quasars just under one million.   

An underrepresented class of absorption line systems in recent searches is the one of Lyman limit systems (LLSs), which are pockets of gas that are optically-thick  to ionizing radiation. Astrophysically, LLSs are thought to trace the denser and partially-ionized regions of the IGM and the CGM \citep[e.g.][]{faucher2011,fumagalli2011,voort2012}. For this reason, they are believed to be excellent sign posts of gas flows near galaxies, and in particular of the cold accretion of gas onto high redshift galaxies \citep[e.g.][]{fumagalli2013,fumagalli2016b}.

\begin{figure*}
    \centering
    \begin{tabular}{cc}
    \includegraphics[scale=0.55]{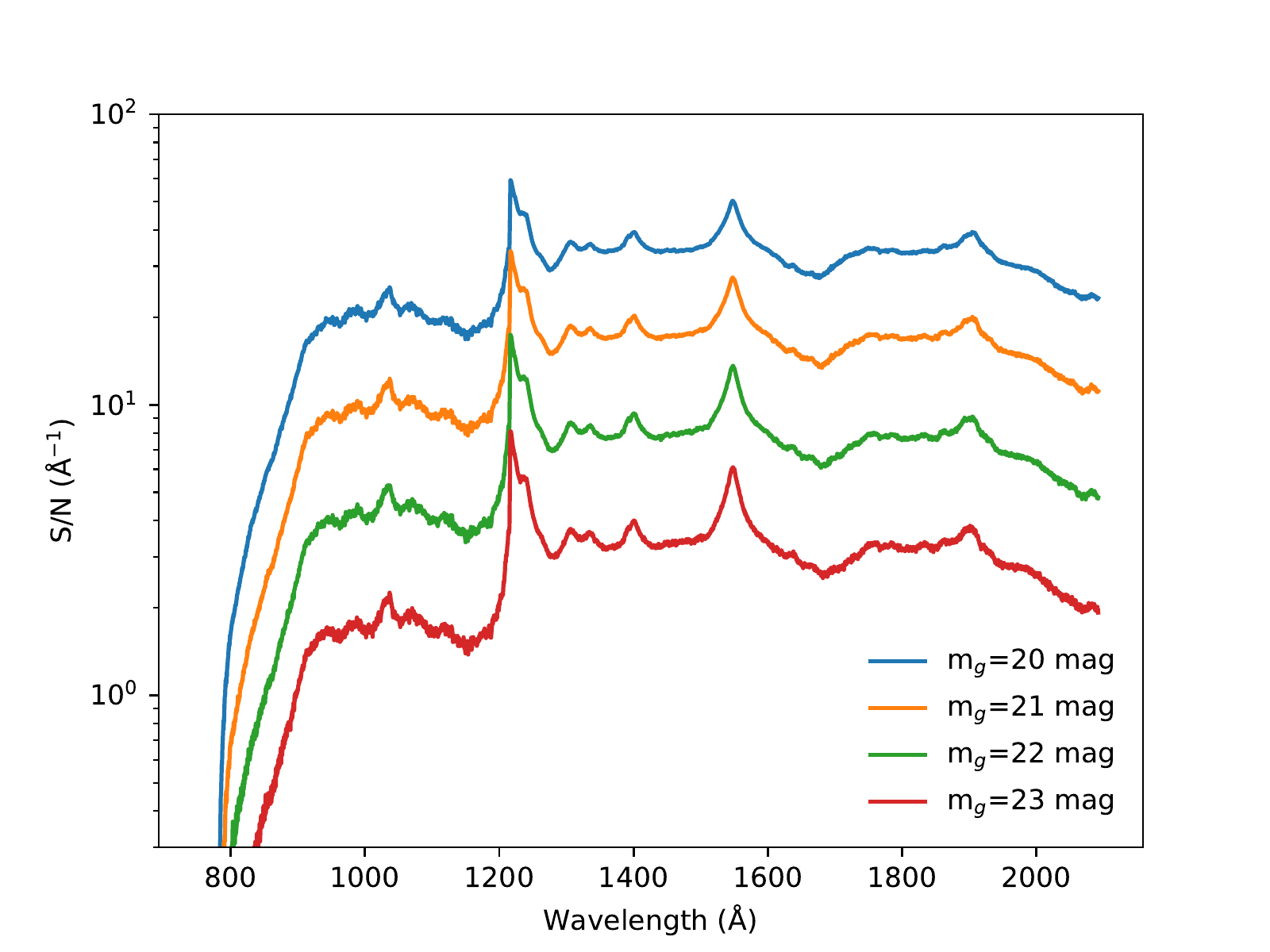}&  
    \includegraphics[scale=0.55]{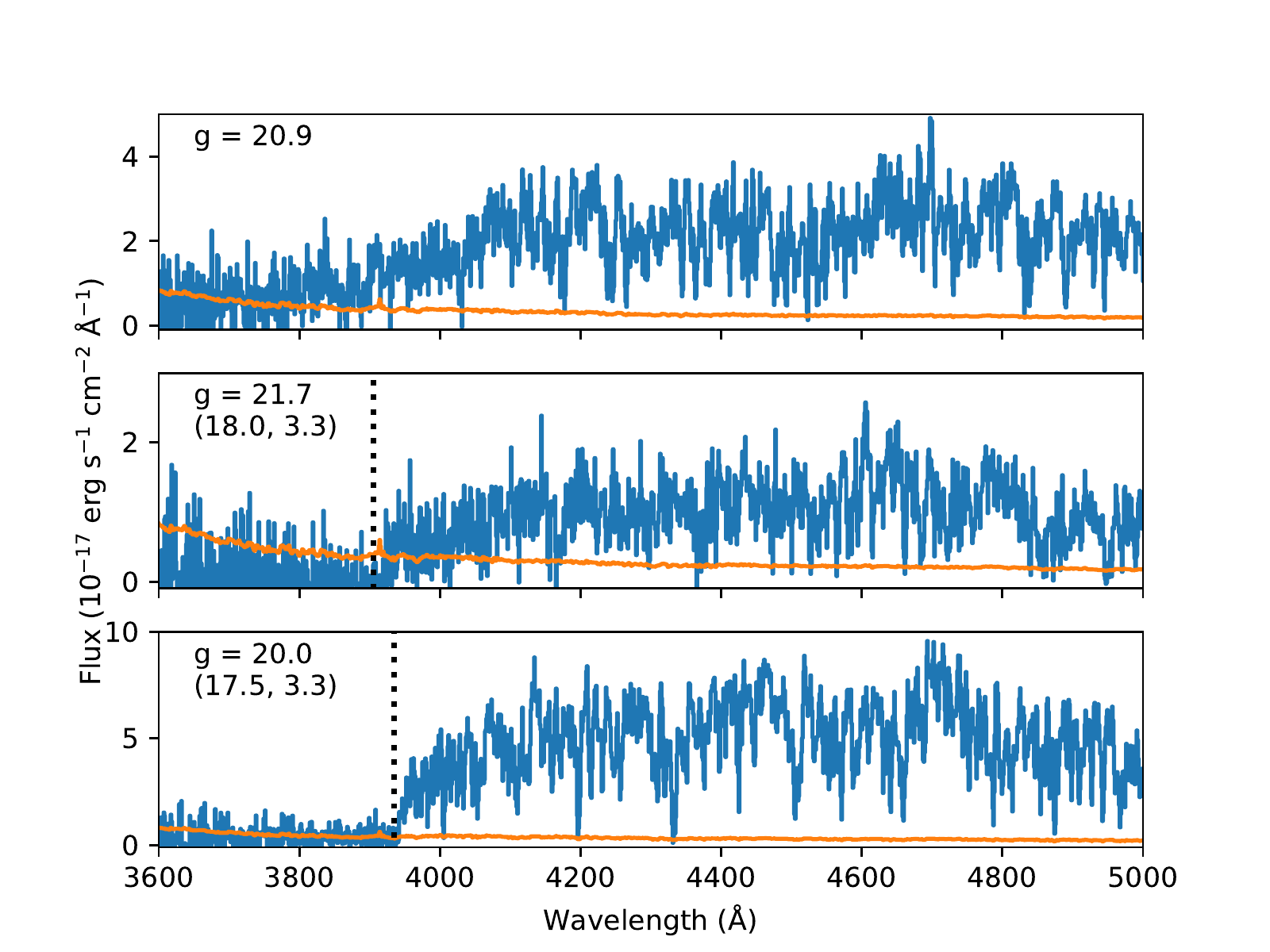}
    \end{tabular}
    \caption{Left: Median $S/N$ of mock spectra as a function of rest-frame wavelength computed for different magnitudes (as color-coded in the legend, in a 0.1~mag interval) and assuming an exposure time of $t_{\rm exp}=4000~$s. Right: Example of mock spectra (in blue the flux and in orange the associated error) for quasars without (top panel) and with LLSs at different $S/N$ (middle and bottom panels). When LLSs are present, we annotate next to the spectrum the value of the \ion{H}{I} column density and the redshift of the absorber. The location of the Lyman limit is marked by a black vertical dotted line.}
    \label{fig:mocksn}
\end{figure*}

LLSs with hydrogen column density $N_{\rm HI} \gtrsim 10^{17}~\rm cm^{-2}$ exhibit saturated Ly$\alpha$ and Ly$\beta$ lines but lack the characteristic wings of a damped profile when $N_{\rm HI} \lesssim 10^{19}~\rm cm^{-2}$. In this column density interval, therefore, they are not readily distinguishable from the lower column density Ly$\alpha$ forest lines based purely on the first few lines of the Lyman series. Due to this reason, algorithms developed for the identification of DLAs are not immediately applicable to the search of LLSs. 
However, at sufficient high redshift ($z\gtrsim 3$), the characteristic flux discontinuity at the Lyman limit (912~\AA\ in the system's rest frame) enters the optical wavelength range. This feature can therefore be targeted by machine learning algorithms with the goal of recognising and classifying this class of absorbers (see Fig.~\ref{fig:mocksn}). 

In this paper, we therefore investigate the possibility of identifying LLSs in large spectroscopic surveys using machine learning. First, we develop a pipeline using mock spectra that are representative of the data quality of DESI and WEAVE (Sec.~\ref{sec:mockprep}-\ref{sec:mockanalysis}). Next, we apply machine learning to identify and compile a catalogue of LLSs using SDSS spectra (Sec.~\ref{sec:sdssanalysis}) and we study the number per unit redshift of $z\approx 3.5$ LLSs (Sec.~\ref{sec:llsprop}). Finally, we test the performance of random forest in the classification of BALs (Sec.~\ref{sec:mockbal}). A summary and conclusions follow in Sec.~\ref{sec:concl}.

\section{Preparation of mock data}\label{sec:mockprep}

In our analysis, we employ two types of data sets: observations of quasar spectra from SDSS (Data Release DR16; Sec.~\ref{sec:sdssanalysis}) and mock libraries of high-redshift quasars (Sec.~\ref{sec:mockanalysis}), assembled in the following way. 
We generate two sets of mock observations of $\approx 180,000$ high-redshift quasars using the DESI survey as reference, including a ``BAL set" which contains BAL quasars and a ``LLS set" containing absorption line systems with $N_{\rm HI} \gtrsim 10^{17}~\rm cm^{-2}$. Mocks are constructed with a two-step process. 

In the first step, we generate sets of idealized noise-free spectra using the quasar template generator {\sc desisim.templates.QSO} in the {\sc desisim} package\footnote{\url{https://github.com/desihub}}. 
Given an interval of redshifts and magnitudes, 
which for the purpose of our analysis we assume to be flat in the interval $m=(17.0,22.7)$ and $z=(3.5,3.7)$\footnote{The choice of redshift interval is dictated by the need to cover the Lyman limit of intervening systems at $z\gtrsim 3$.}, the code generates quasar spectra from a principal component analysis, with  coefficients derived from real SDSS observations. These spectra are then populated with a realization of the Ly$\alpha$ forest using the 1D power-spectrum method described in \citet{macdonald2006}, which reproduces a realistic-looking transmitted flux.  

The flux bluewards of 912~\AA\ in the quasar rest-frame is further attenuated by an exponential function that mimics the absorption of an unresolved population of strong absorbers, with column densities $\gtrsim 10^{15}~\rm cm^{-2}$ \citep[e.g.][]{prochaska2009,prochaska2010,fumagalli2013}.
The scale-length of the exponential attenuation is modulated by the mean free path of ionising photons, $\lambda_{\rm mfp}$, described by a distribution with mean  
\begin{equation}
\lambda_{\rm mfp}= 37\left(\frac{1 + z}{5}\right)^{-5.4}{\rm Mpc}
\end{equation}
and a $15\%$ dispersion \citep[see][]{prochaska2009,fumagalli2013}.

 Starting with this set, spectra in the BAL set are then populated with BALs, including \ion{H}{I} and \ion{C}{IV} features, using a set of pre-computed templates \citep[see][]{guo2019}. The probability to host a BAL is set to $P_{\rm bal} = 0.5$, much above the canonical value for a mock sample which is representative of the real Universe. For the purpose of our pipeline development, however, this choice ensures that a good number of BAL sightlines are available for training and testing of the identification procedure. 

For the LLS set, we include instead a population of LLSs with $N_{\rm HI} \ge 10^{17.2}~\rm cm^{-2}$, modelling the full Lyman series and the Lyman limit. LLSs, which in our working definition also include DLAs, are injected at random, with a probability defined by $P_{\rm lls} = p_{\rm lls} \ell(z) \Delta z $, where $\ell(z)$ is the integral of the column density distribution function from \citet{prochaska2014} and $\Delta z$ is the path length defined by the quasar redshift and the redshift for which a Lyman limit would fall at the bluest observable wavelength in a spectrum. Here, $p_{\rm lls} = 0.3$ to ensure that a significant number of spectra do not contain LLSs. Again, while not strictly representative of the real number per unit redshift of LLSs, this choice produces a balanced mix of sightlines with and without LLSs for training and for testing the classification. Column densities are drawn at random from the column density distribution function.  
As our primary goal is not to reproduce a mock set representative of the real Universe but simply to develop the formalism necessary for the recovery of LLSs in individual spectra, we do not account for a correlation between LLSs and the Ly$\alpha$ forest, and do not include metal lines.

If a sightline contains a LLS with optical depth $\tau < 2$ at the Lyman limit, we further allow for the presence of a second LLS, by repeating the procedure above and setting the path length to the range defined by the first LLS and the blue edge of the spectrum. 
Similarly, we allow for additional lower-redshift DLAs, again using the procedure above for the relevant column density range ($N_{\rm HI} \ge 10^{20.3}~\rm cm^{-2}$)
and by setting the path length to the redshift range of the last LLS and the redshift where Ly$\alpha$ falls at the bluest end of the spectrum. 
The injection of these additional LLSs and DLAs allows us to capture the additional sources of large-scale absorption that alter the shape of the quasar continuum.

During the second step of the mock preparation, these noise-free spectra are fed into the DESI {\sc quickspectra} tool\footnote{\url{https://github.com/desihub/desisim}} that, albeit working only in 1D, generates realistic mock observations accounting for the relevant instrument features and mimicking the typical observing conditions of the DESI survey \citep{kirkby2016}. The final product is a set of spectra with resolution and noise properties that are characteristic for DESI data, although we expect that these mock data reflect more generally the quality of observations at 4m telescopes, e.g. from the WEAVE survey. 

For both the LLS and BAL sets, we generate three ``surveys"  starting from the same input mock spectra: a shallow, a medium, and a deep survey simulated, respectively, with an exposure time of 1000~s, 4000~s and 16000~s (i.e. in increments of signal to noise ratio  $S/N=2$). Median $S/N$ ratios as a function of rest-frame wavelength in bins of quasar magnitudes are shown in the left hand-side panel of  Fig.~\ref{fig:mocksn} for the case of $t_{\rm exp} = 4000~$s in the BAL set.
The right hand-side panel shows instead examples of mock spectra with and without LLSs at different $S/N$.

\section{Analysis of mock spectra}\label{sec:mockanalysis}

\subsection{Classification of LLSs}\label{sec:mockllsclass}

For the classification of LLSs, we rely on their most distinctive feature, 
which is the flux break arising from the hydrogen bound-free transition at 912~\AA\ in the rest-frame of the system (Fig.~\ref{fig:mocksn}): $f(\lambda) = \bar{f}(\lambda)\exp[-\tau(\lambda)]$, where
$\tau(\lambda)\approx \hat{N}_{HI}\hat{\lambda}^{3}$
for $\lambda < 912$~\AA, with $\hat{N}_{HI}= N_{HI}/10^{17.2}~\rm cm^{-2}$ and $\hat{\lambda}=\lambda/912~$\AA.
Here, $\bar{f}(\lambda)$ is the intrinsic spectrum, i.e. the spectrum impingent on the intervening LLSs.
While for DLAs or high column density LLSs with $N_{\rm HI} \gtrsim 10^{19}~\rm cm^{-2}$ the shape of the first Lyman series lines adds key information on the column density, the Ly$\alpha$ and Ly$\beta$ lines in lower-column density LLSs with $N_{\rm HI} \lesssim 10^{19}~\rm cm^{-2}$ are fully saturated and without damping profiles, making them indistinguishable from the rest of the lower-column density Ly$\alpha$ forest. The depth of the 912~\AA\ break is instead sensitive to the column density, or to its lower limit for $\tau \gg 2$ LLSs. This is the feature we target for our classification. 

Requiring the presence of the redshifted Lyman limit within the spectral range, however, introduces a stringent lower limit on the redshift of absorption systems, and hence of useful quasars, that can be analysed with this technique. Indeed, to ensure that the Lyman limit appears above $3700~$\AA\ where the end-to-end throughput of common spectrographs is still sufficiently high, LLSs have to be above $z\approx 3.05$, making $z\gtrsim 3.5$ the ideal sightlines for this search. As the majority of the spectra in large surveys are targeted to optimally cover the Ly$\alpha$ forest at $z\lesssim 3$, this redshift requirement results in a non-negligible reduction in sample size. 

For our classification, we use the
random forest classifier \citep{breiman2001}
as implemented in the {\sc RandomForestClassifier} method  within \scikit\ \citep{scikit-learn}.  We first homogenize the spectra by normalising the quasar continuum to remove the intrinsic flux differences among quasars. In principle, this could be accomplished by fully modelling the underlying quasar continuum (see e.g. Sect.~\ref{sec:mockbal}). It is however notoriously difficult to reproduce the shape of the quasar spectrum at $\lesssim 912~$\AA\ in the quasar restframe, given not only the presence of individual LLSs, but also the variation from one sightline to another in the mean IGM attenuation. Nevertheless, over a narrow range of wavelength ($\approx 800-900$~\AA\ in the quasar rest frame) the continuum is generally characterized by a power law combined with an exponential profile, the scale-length of which is defined by the mean free path \citep{prochaska2009}.  Thanks to the intrinsic homogeneity of both the quasar spectral energy distribution and the mean free path in a given redshift bin, we can afford to omit a full continuum normalization of the spectrum. Instead, we simply apply a constant normalization factor computed between $915-940$~\AA. During the training of the random forest classifier, residual variation in the shape of the spectrum will be learned as an irrelevant feature, while significant departures due to intervening LLSs will be used as the main feature to classify.

\begin{figure}
    \centering
     \includegraphics[scale=0.55]{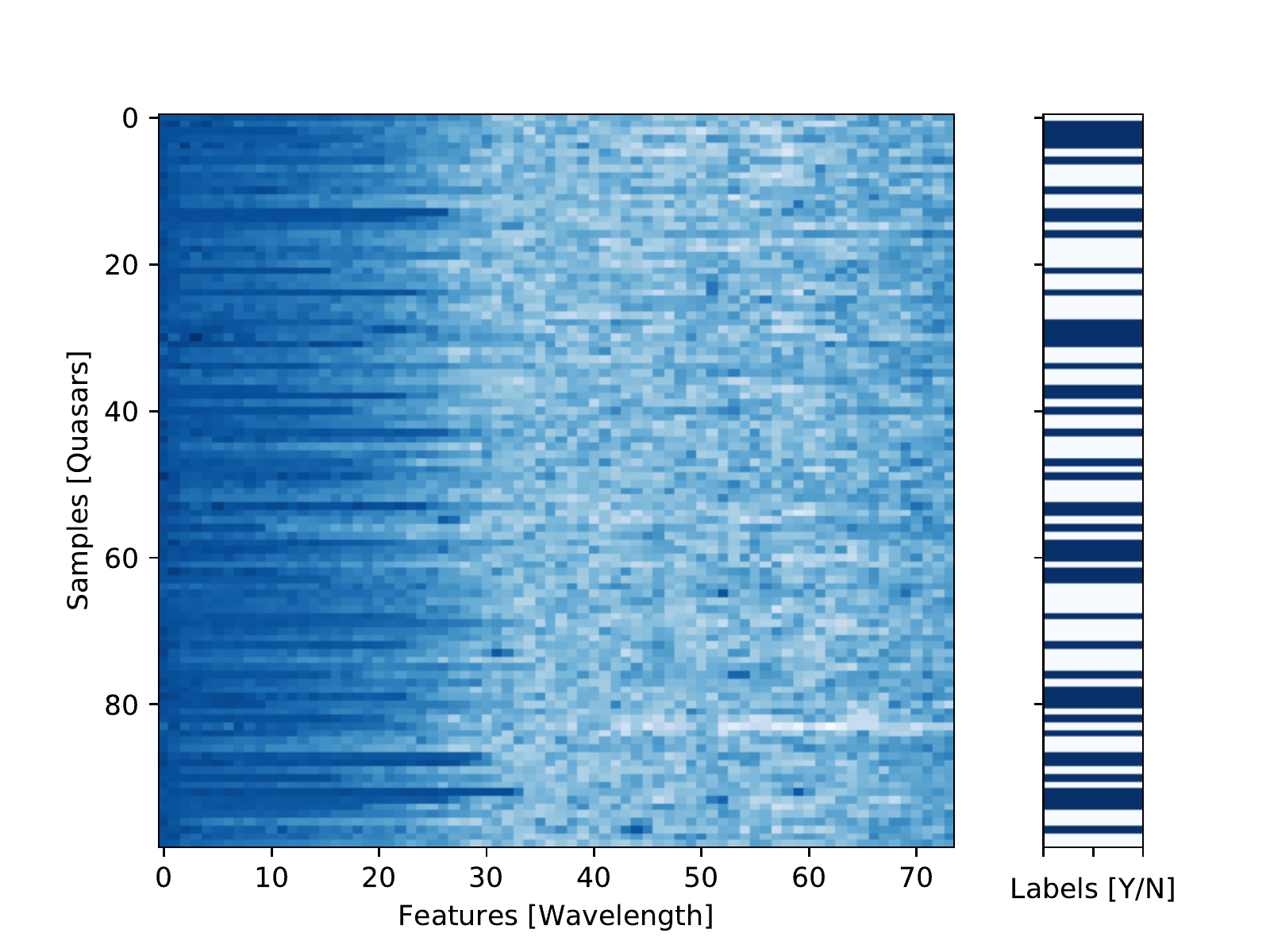} 
    \caption{Vertical stack of 100 mock quasar spectra from the LLS set with $t_{\rm exp}=4000$~s. Spectra of individual quasars are stacked along the vertical direction, with the wavelength (in units of pixels) increasing along the x-axis from left to right. The color map encodes the normalized flux in the range (0,1) where 0 is dark blue and 1 is white. The dark stripes visible leftwards of index $\approx 30$ correspond to the flux decrement of $\tau \gtrsim 2$ LLSs at their Lyman limit. The color bar to the right indicates which sightlines contain LLSs (blue) and which do not (white).}
    \label{fig:llsfeatures}
\end{figure}

\begin{figure*}
    \centering
    \begin{tabular}{c|c}
    \includegraphics[scale=0.55]{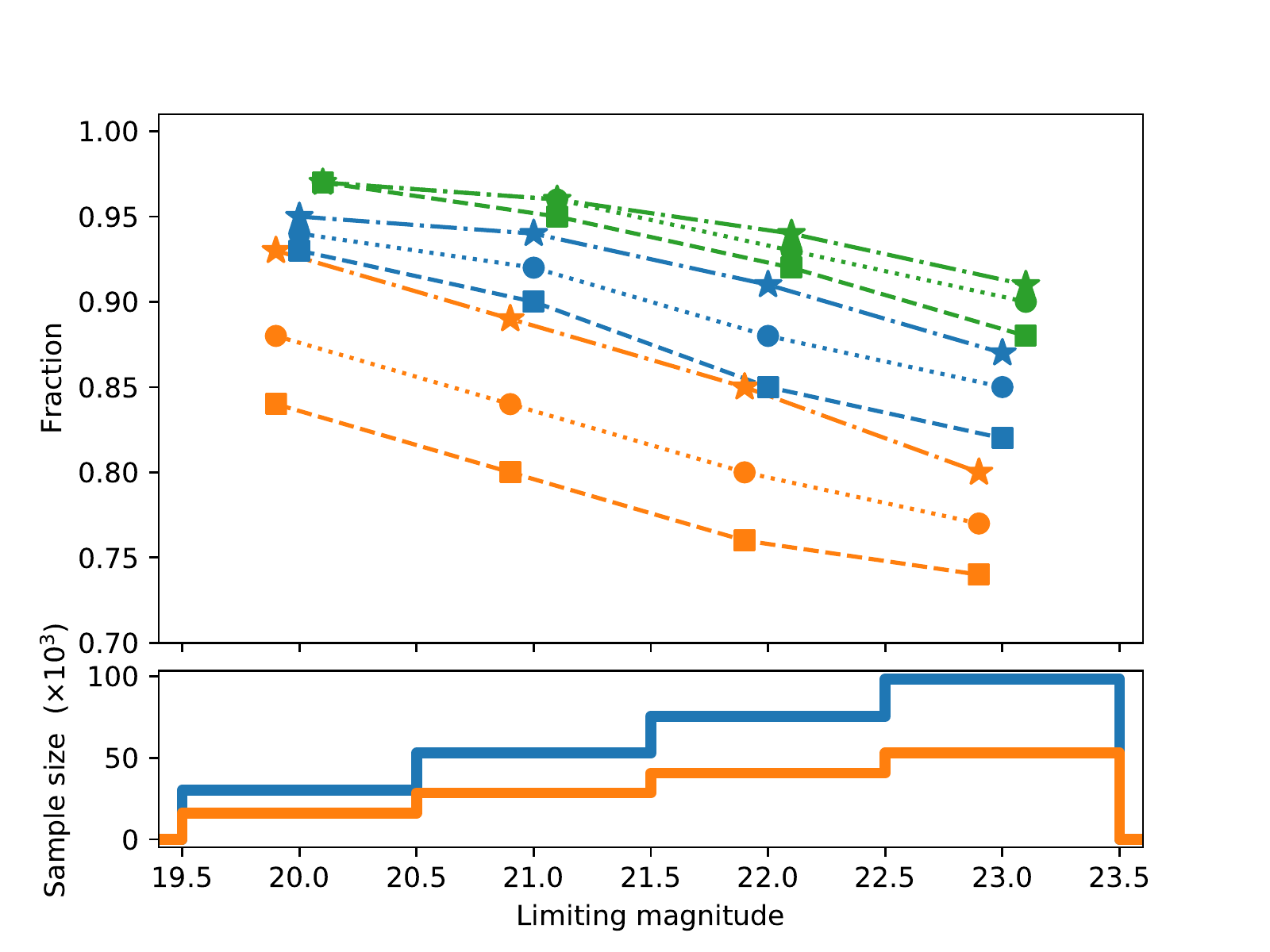} & 
    \includegraphics[scale=0.55]{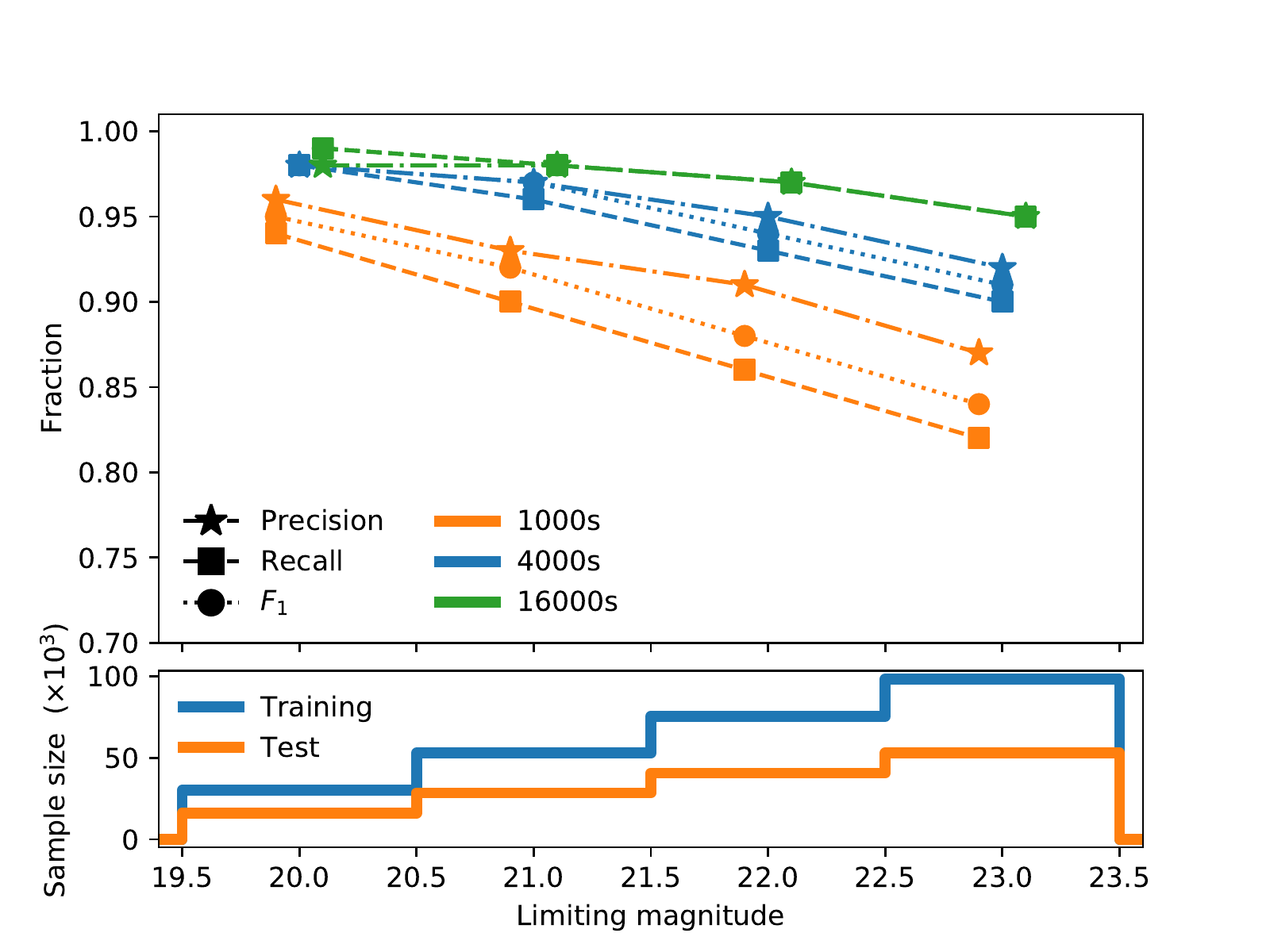} 
    \end{tabular}
    \caption{Summary of the statistics associated with the classification of LLSs in mock data. The top panels show the precision (stars and dash-dotted lines), recall (squares and dashed lines), and $F_1$ score (circles and dotted lines) as a function of limiting magnitude ($g$ band). Values for three exposure times are shown, as color-coded in the legend. The bottom panels show the number of quasars included in the training set (blue) and test set (orange). The left and right panels are for the native and $20\times$-binned spectra, respectively. The completeness and purity of the LLS sample depend on the spectral $S/N$, making binned data preferred for this analysis.}
    \label{fig:llsstats}
\end{figure*}

\begin{figure}
    \centering
    \includegraphics[scale=0.55]{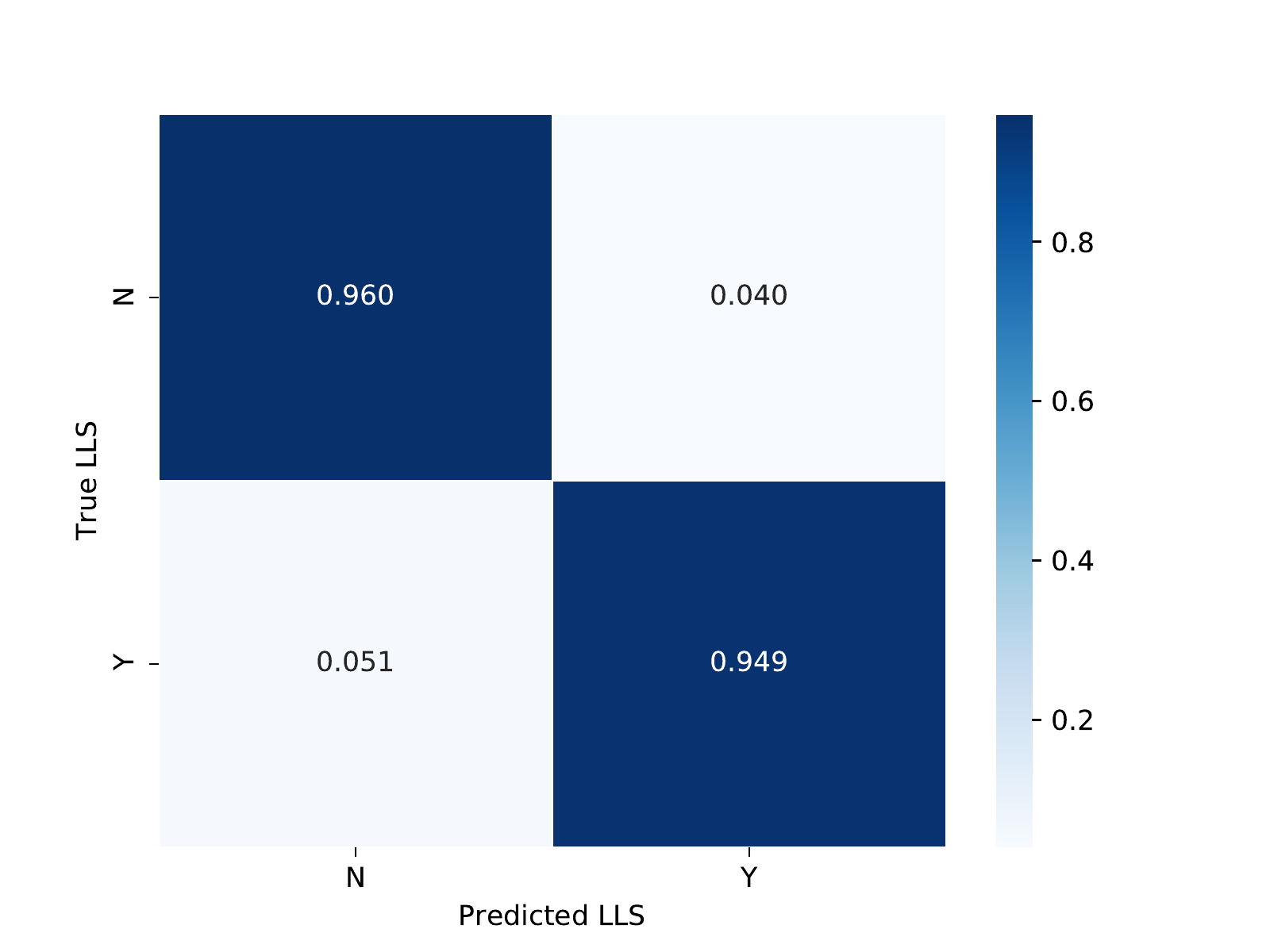} 
    \caption{Confusion matrix for the classification of quasars with $t_{\rm exp}=4000~\rm s$ and $m_g \le 21.5~\rm mag$. }
    \label{fig:cmatrix}
\end{figure}

\begin{figure*}
    \centering
    \begin{tabular}{c|c}
    \includegraphics[scale=0.55]{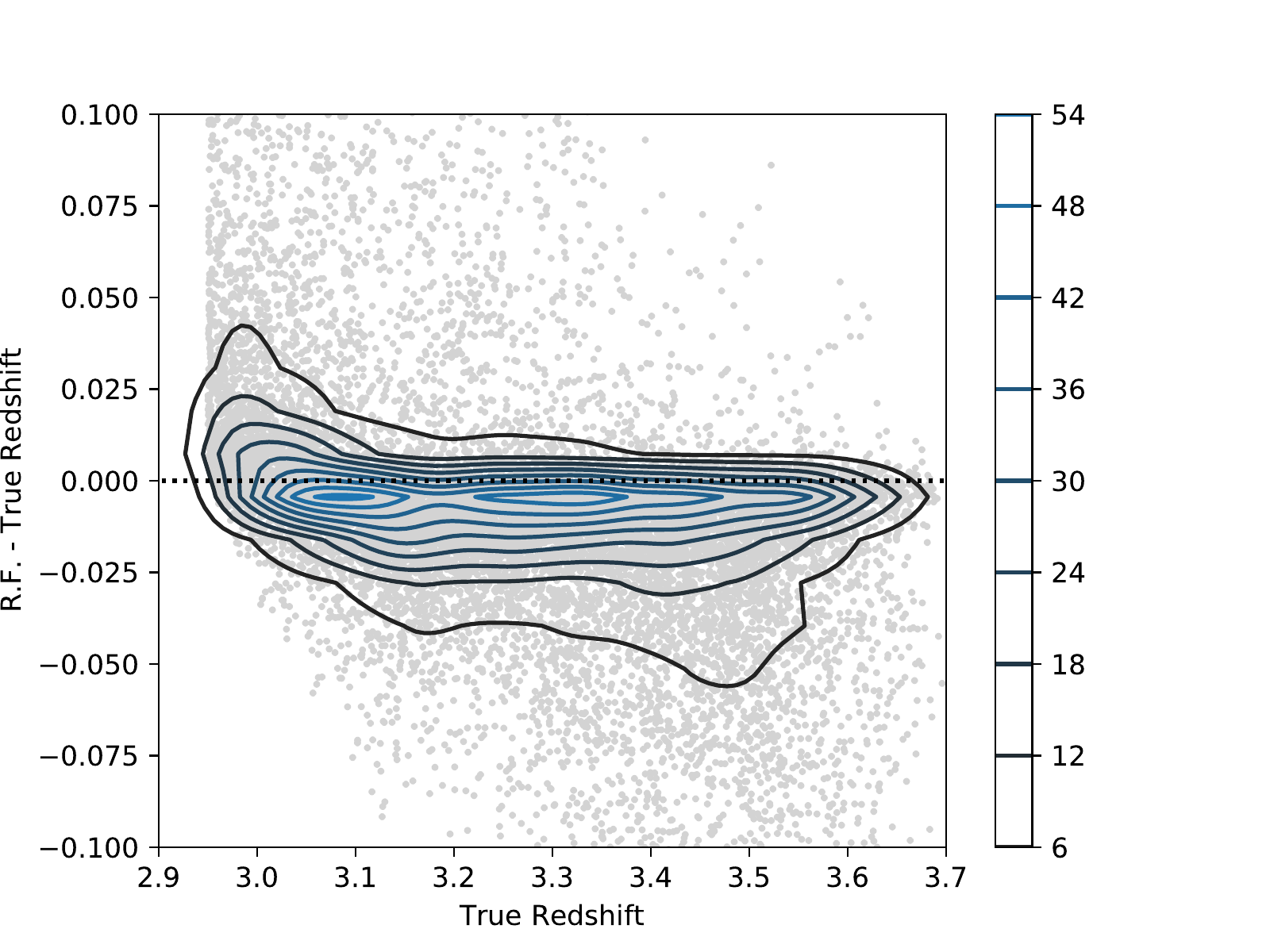} &
    \includegraphics[scale=0.55]{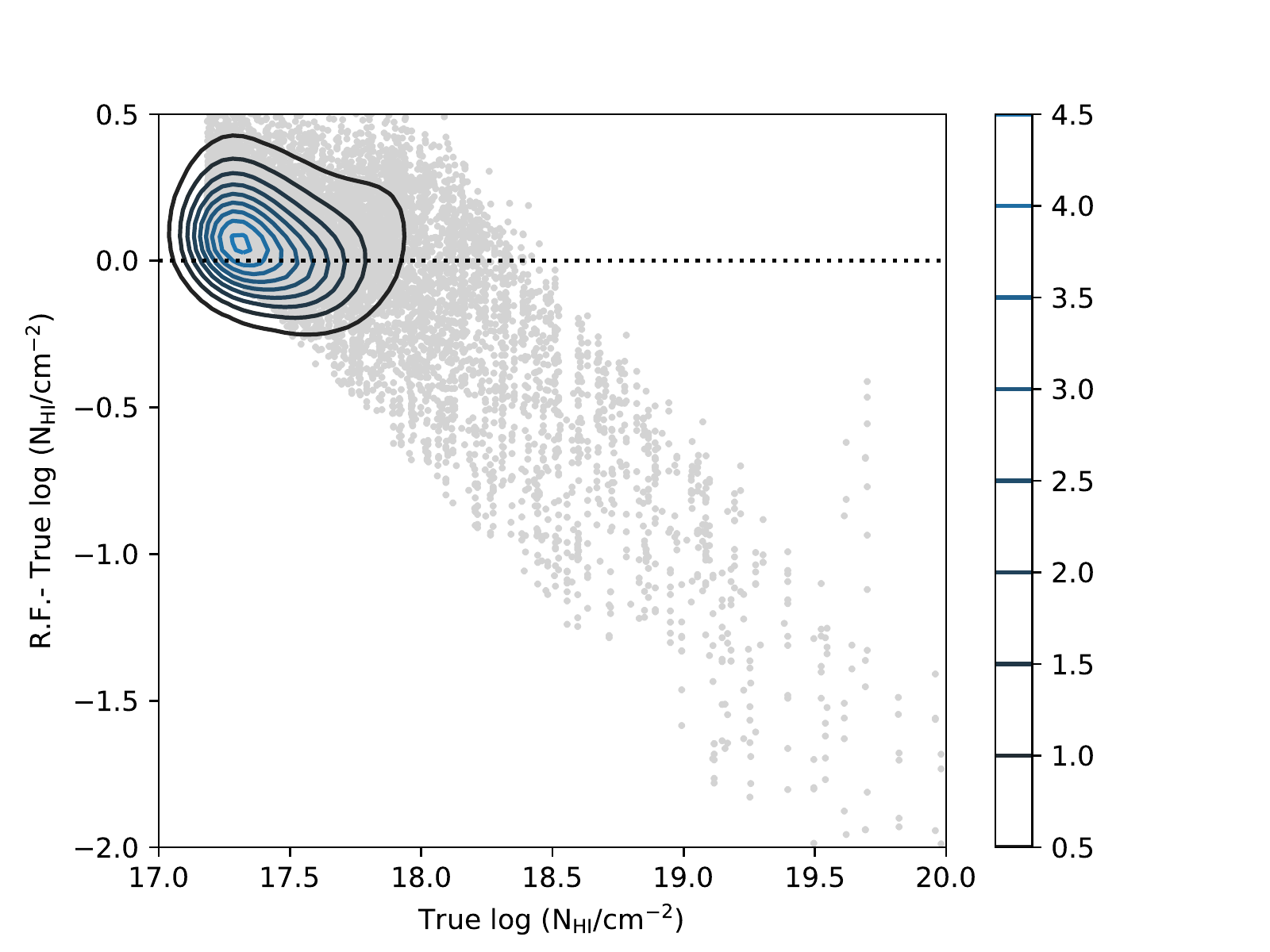}  
    \end{tabular}
    \caption{Comparison between input and recovered redshift (left)
    and 
    \HI\ column density (right)  for LLSs in mock spectra of $m_{\rm g}< 21.5$ quasars with smoothing factor $20\times$ and $t_{\rm exp}=4000~$s. Kernel density estimates are shown on the top of the scatter plots. While redshifts are accurately recovered for most of the sample, the \HI\ column density for systems with $\tau \gtrsim 2$ is difficult to constrain based on the flux decrement at the Lyman limit alone, especially in noisy data.}
    \label{fig:llsprop}
\end{figure*}

Following the continuum normalization, we arrange the spectra in a $n_{\rm qso} \times n_{\rm pix}$ matrix in the observed frame, as shown in Fig.~\ref{fig:llsfeatures}. Ideally, we would align spectra in the quasar rest frame, but given that the features of interest lie in the bluest portion of the spectrum, differences in redshifts would imply that we would lose part of the pixels of interest for quasars at the lower end of the redshift distribution. However, the fact that we are working in a 
narrow redshift window, $z=(3.5,3.7)$, means that the spectra are already ``quasi-aligned" in the rest frame. Furthermore, there are no prominent emission lines in the region of interest ($<1100~$\AA\ in the rest frame at the median redshift of the quasar sample, which encompasses the Lyman limit of all quasars), so moderate shifts in wavelengths will not significantly affect our classification. 

After this preparatory step, we proceed to train  the {\sc RandomForestClassifier} method starting with the optimization of the parameters that control the algorithm. For this, we use 65\% of the mock sample simulated with an exposure time of 4000~s, after applying a magnitude cut at $m_{\rm g} < 21.5~\rm mag$ to exclude the lowest $S/N$ data. Quasars with at least one LLS are tagged as containing absorbers, and no distinction is made for the case of one versus multiple LLSs. Using a five-fold cross-validation with $F_1$ score as metric of performance, we find an optimal classification ($F_1 \approx 0.90$) for {\tt n\_estimators=500}, {\tt criterion=`entropy'}, {\tt min\_samples\_split=2}, and {\tt max\_features=400}. As we are classifying a flux drop over several tens of angstroms, it is not surprising that the best performances are found for a large number of maximum features. We also find that the classification is not very sensitive to the exact choice of parameters, with $F_1 \approx 0.88-0.90$ for reasonable variations.

Following this tuning operation, we apply the classifier to mock samples  with different exposure times and with different magnitude cuts. For every combination, we first train the algorithm on 65\% of the dataset, and then apply the classification to the remaining portion that has never been seen by the classifier. Results are summarized in the left panel of Fig.~\ref{fig:llsstats}. The performance of the random forest classifier is generally good, with typical completeness (recall) and purity (precision) in the range of $\approx 80-95\%$. It is also evident that the algorithm is particularly sensitive to the spectrum $S/N$, with a clear dependence on the magnitude limit and on the simulated integration time. 

Due to the strong sensitivity of the $S/N$, and given that the feature we aim to classify is not too sharp in wavelength, we explore the performance of the search following re-binning of the data. The right panel of Fig~\ref{fig:llsstats} shows the result of the classification for spectra that have been rebinned by a factor of 20 (corresponding to $\approx 4$\AA). The maximum number of features used for the classification is scaled accordingly. 
Once the data have been rebinned, we see a marked improvement in the recovery of LLSs, with precision and recall consistently above $90\%$ for $t_{\rm exp} \ge 4000~$s. 
The confusion matrix for quasars with $m_g<21.5~\rm mag$ is shown in Fig.~\ref{fig:cmatrix}. Doubling the size of the bins leads to further but rather marginal improvement in the classification performance. 

Examining the mocks that are misclassified, we find that the main reason for incompleteness is the redshift of the LLSs. Indeed, we find that $>50\%$ of the missed LLSs lie at $z<3$, and $\approx 80\%$ of them lie at $z<3.1$. This is not surprising, as the lowest redshift LLSs will present a Lyman limit at the edge of the spectral range, where the $S/N$ is intrinsically low (see Fig.~\ref{fig:mocksn}) and where only a handful of pixels are covered bluewards of the system's Lyman limit. Regarding the sightlines that are incorrectly classified as hosting a LLSs, we do not identify a clear property to which we can attribute the incorrect classification, although we note that spectra appear to consistently lie below the mean spectrum of the entire sample in the region $<912~$\AA\ in the quasar rest frame. These sightlines, with intrinsically redder quasar spectra or lower mean free path, may indeed more easily be mistaken at modest $S/N$ as containing LLSs.

\begin{figure}
    \centering
    \includegraphics[scale=0.53]{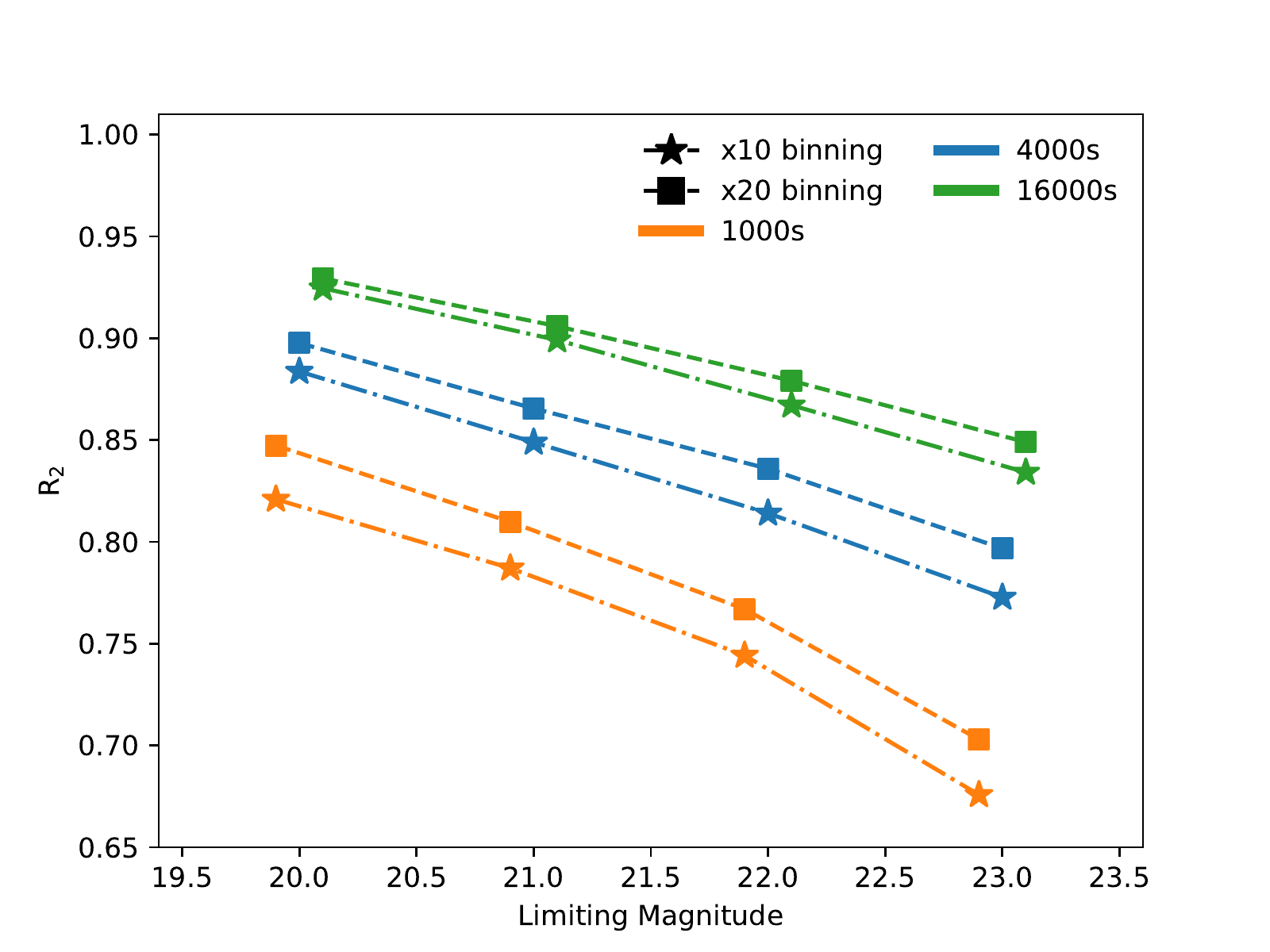} 
    \caption{$R_2$ coefficients for the recovery of LLS redshifts using the random forest regressor as a function of limiting magnitude ($g$ band) for different choices of binning (stars and squares for $\times 10$ and $\times 20$ binning, respectively) and exposure times (color-coded as in the legend). Redshifts are generally recovered successfully, particularly at higher $S/N$.}
    \label{fig:llspropstats}
\end{figure}

\subsection{Measuring redshift and column density}

Having searched for LLSs in the mock sample, we next examine the possibility of recovering physical parameters such as redshift and column density from the spectra. To this end, we employ the {\sc RandomForestRegressor} in \scikit, which we use to fit for a value of redshift and column density following a procedure similar to the one for the identification of LLSs.

At first, we prepare the data as in Sect.~\ref{sec:mockllsclass}, restricting only to mock sightlines which host LLSs (a property that becomes known once the random forest classification has been applied). Following the pre-processing of the data as in the previous section, we define a training set for the regressor using $2/3$ of the data, and test the code performance on the remaining $1/3$ of the sample. In presence of multiple LLSs, we focus on the strongest absorber in the line of sight, similarly to the procedure that human classifiers would follow (see the next section).  

An example of the recovery of both the \HI\ column densities and the redshifts using 
{\tt n\_estimators=500}, {\tt criterion=`mse'}, {\tt min\_samples\_split=2}, and {\tt max\_features=`auto'} is shown in Fig.~\ref{fig:llsprop}, for the test set of $m_{\rm g} < 21.5$~mag mock spectra which are simulated with $t_{\rm exp} =4000$s  and are re-binned by $20\times$ the native pixel size. As evident from this figure, redshifts (right panel) are correctly recovered across the entire range with only a mild bias towards lower values for $z\gtrsim 3.2$. Also evident is a population of outliers close to $z\approx 3$, for which the random forest regression predicts less accurate values that are overestimated. We attribute this effect to the same difficulty of correctly identifying LLSs  close to the edge of the spectral range, as noted in the previous section. Specifically, we hypothesize that the lower $S/N$ of these spectra makes them appear more absorbed than they really are, causing an overestimation in the random forest redshifts. There are also some outliers around $z\approx 3.4-3.6$, where the random forest redshift is underestimated compared to the true value. Possible explanations for this discrepancy are features in the high-order Lyman series and/or in the quasar continuum that skew the redshift determination. 

Overall, however, the performance of the random forest regression is good, as quantified by a coefficient of determination $R_2\approx 0.85$. As in the case of the classification, we note that the performance of the regression method is a function of the spectral $S/N$, with the best results obtained for binned data of brighter quasars or quasars observed for longer time (see Fig.~\ref{fig:llspropstats}). 

Regarding the determination of the column density (left panel of Fig.~\ref{fig:llsprop}), we see instead a reasonable recovery only up to $N_{\rm HI}\approx 10^{17.5}~\rm cm^{-2}$, at which point the recovered values start deviating from the true value. This is not surprising, as for  $N_{\rm HI}\gtrsim 10^{17.5}~\rm cm^{-2}$ (i.e. $\tau \gtrsim 2$) the flux decrements at the Lyman limit reaches approximately its saturated value, making difficult to distinguish subtle differences in the flux decrements for higher column densities in the absence of other information (e.g. the shape of the Lyman series). The outcome of this analysis is therefore in line with the expected performance given the information content of the spectra. A more accurate determination of column densities requires a detailed model of the entire spectrum,
as Lyman series lines are expected to carry additional information to pin down the value of $N_{\rm HI}$. While the recovery of  the column density will be successful for $N_{\rm HI} \gtrsim 10^{19}~\rm cm^{-2}$,
it may be still be difficult to achieve a reliable determination for most LLSs between $10^{17.5}~\rm cm^{-2} \lesssim N_{\rm HI} \lesssim 10^{19}~\rm cm^{-2}$  where Lyman series lines are in the saturated portion of the curve of growth \citep[e.g.][]{prochaska2015}.

\section{Searching LLS\texorpdfstring{\MakeLowercase{s}}{s} in SDSS/DR16}\label{sec:sdssanalysis}

\subsection{Preparation of the training set}

We apply the pipeline developed above to $\approx 10,000$ SDSS DR16 spectra observed with the BOSS spectrograph \citep{Dawson2013} and classified as quasars by the automatic pipeline \citep{Bolton2012}.
In the full sample, we allow for duplication of spectra for quasars that have been observed more than once.  
For this proof-of-concept analysis, we elect the redshift range $3.5 \le z \le 4.0$, which offers optimal coverage of the Lyman limit down to $z\approx 3$ and includes a sufficiently large number of quasars. 
Due to expected differences 
between the mock data set used above and the BOSS data, we re-train the random forest classifier using a training set extracted from the observed spectra, using however the same methodology that we have validated in the previous section with mocks.  

To compile this training set, we visually inspect 2010 spectra using a custom-made GUI that allows the classification of quasars in three classes: i) clear sightlines, in which we do not detect a LLS; ii) sightlines with at least one LLS, for which we record the redshift based on the location of the Lyman limit and a crude estimate of the column density based  on the depth of flux decrement at the Lyman limit; iii) ambiguous sightlines for which the noise prevents a clear classification in the former two classes. To avoid the ambiguity arising from the presence of partial LLSs in moderate $S/N$ data, we limit our classification only to $\tau \gtrsim 2$ LLSs, which are strong enough to produce a clear and (nearly) saturated absorption at the Lyman limit.
Furthermore, as the data quality is significantly degraded for $\lambda \approx 3750~$\AA, we classify only systems with LLSs at $z\gtrsim 3.11$, marking as ``clear" the (very few) sightlines for which LLSs are recognizable at lower redshift. Thus, in our analysis we truncate the data, including only wavelengths above $3750~$\AA.  

During the preparation of the training set, we also encountered a small number ($65/2010$) of spectra that have been incorrectly classified by the SDSS pipeline as quasars (mostly blue stars, quasars at lower redshift or artefacts with bumps mimicking emission lines). We add these spectra to a further ``non quasar" class.  Before proceeding to the classification of LLSs, we clean the remaining portion of the catalogue by applying the random forest classifier
to spectra that are binned by a factor of 40 of the original resolution, aligned in the quasar rest frame, and normalized at 1450~\AA. Using a five-fold cross-validation on the training set, we tune the classifier to yield a maximal recall of objects that are not quasars, with the goal of rejecting as many spurious sources as possible (at the expense of sample size). 

Due to the very limited size of the class we aim to identify, it is not surprising that the algorithm achieves a recall of only $70\%$ for the true class (i.e. non quasars) albeit with a precision of $95\%$. Indeed, in the remaining portion of the catalogue, the classifier identifies $\approx 1.6\%$ of spurious sources which we subsequently confirm as non quasars through visual inspection. Due to the lower fraction of non quasar automatically identified compared to that of the training set ($1.6\%$ vs $3.2\%$), it is likely that a small number ($\approx 1\%$) of sources included in the final catalogue is in fact not a quasar. Such a small fraction will not affect the usefulness of the catalogue as a whole.

\begin{figure}
    \centering
    \includegraphics[scale=0.53]{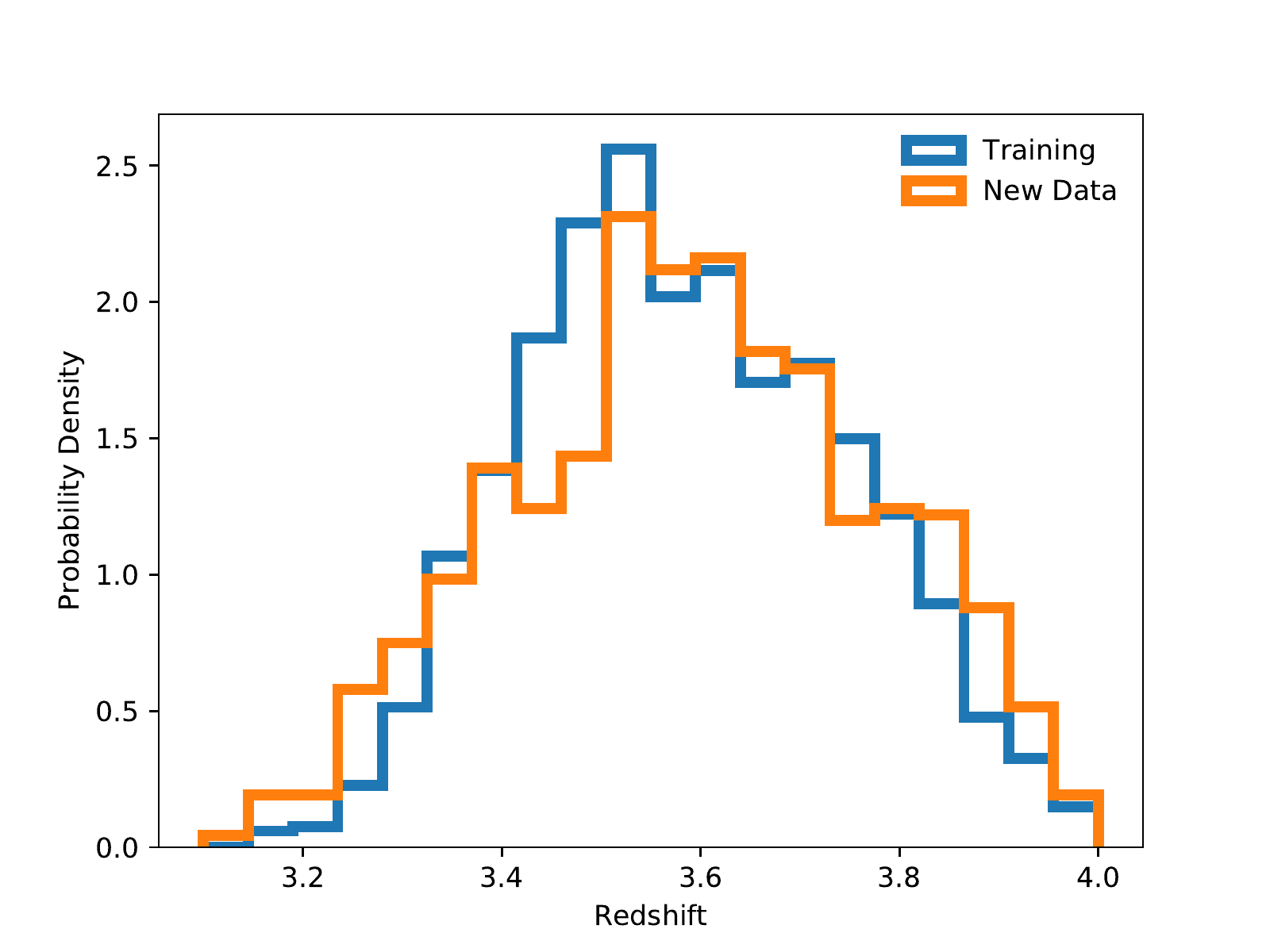} 
    \caption{Redshift distribution of LLSs identified in the training set (blue) and remaining set (orange) of SDSS/DR16 data. A comparable fraction of LLSs is found in both sets, with a similar redshift distribution.}
    \label{fig:llszfit}
\end{figure}

\subsection{Classification and redshift measurement}

With the training set prepared above, we train the random forest classifier on the 2010 BOSS spectra we visually classified, and proceed to apply the classification on the remaining $\approx 7900$ quasar sightlines\footnote{For simplicity, we exclude a very small number of sightlines with wavelengths starting at $\lambda > 3750$~\AA.}.
Throughout this analysis, we interpolate all spectra on a common wavelength grid in the interval $3750-5200$~\AA, with the wavelength range chosen to encompass all LLSs at $z\gtrsim 3.1$ (as described above) up to the quasar Ly$\beta$ emission line.  
The spectra are then continuum normalized below the Lyman limit in the quasar rest frame using a constant factor computed as the 90$^{\rm th}$ percentile between $915-940~$\AA\ in the quasar rest frame.  We further bin the spectra at 20 times the original resolution, based on the results discussed in the previous section. 

Using a five-fold cross-validation, we tune the classifier parameters to yield a maximal $F_1$ score, finding that the best classification is achieved for 
{\tt n\_estimators=1000}, {\tt criterion=`gini'}, {\tt min\_samples\_split=4}, and {\tt max\_features=40}. As noted above, the final classification is only weakly sensitive to variations of these parameters.
For our choice, we find an $F_1$ score of $\approx 0.88$, and a recall and precision for the classification of LLSs of 0.92 and 0.94, respectively. This is broadly in line with the performance of the classification on mock data, given both the smaller size of the training set and the different data quality of DESI and BOSS spectra.
Applied to the remaining sample, the classifier identifies 5580 LLSs, for a total of 6621 LLSs including those identified in the training set. 
Over the entire sample of LLSs, the fraction recovered ($\approx 68\%$) is well matched to the fraction of LLSs present within the training set ($\approx 64\%$). 

Excluding repeated observations (i.e. considering only the ``Science Primary'' quasars in SDSS), we identify 4801 unique LLSs. The presence of repeated observations further allows for testing the repeatability of the classification. Among all the repeated observations, the classifier achieves consistent classification in $\approx 70\%$ of unique quasars. This fraction rises above $80\%$ when restricting to high-quality observations, i.e. those with $S/N>5.5$ at 1150~\AA\ as described below. We note that this test is rather stringent, as in the case of more than two observations for the same quasars, it is sufficient to have a single discordant classification to fail the test. In fact, we find that in most of these cases, the classifier yields consistent classification for the majority of the repeated observations.   

Having identified a sample of $\tau \gtrsim 2$ LLSs,
we proceed by fitting their redshift using a random forest regression with the same choice of parameters used for the mock data (see the previous section). After training the regression on the sample of  1041 LLSs which we visually identified in the training set, we fit for the redshift of the  5580 new LLSs. Fig.~\ref{fig:llszfit} shows that the distribution of redshifts for both the training and the remaining set are consistent with each other, as expected given that the training set is a random subset of the full sample. We do not attempt, however, to fit for the column density, as our selection of $\tau \gtrsim 2$ LLSs puts us in the column density range where there is little to no sensitivity on the column density (see Fig.~\ref{fig:llsprop}). The final classification for the full sample is presented in Table~\ref{tab:llslist}.

\begin{table*}
    \centering
    \begin{tabular}{cccccccccc}
\hline
Name & R.A. & Dec. & z$_{\rm qso}$ & $\rm S/N_{1150}$ & Primary & Train & Class & z$_{\rm lls}$\\
  & (deg) & (deg)&  &   &  &  & \\
\hline
spec-10227-58224-0053& 143.83343& 32.3545430& 3.527&  9.6& Y& Y& 1& 3.518\\
spec-10228-58223-0374& 145.21736& 32.1541970& 3.757&  4.1& Y& Y& 2& 0.000\\
spec-10228-58223-0474& 144.60158& 33.1317990& 3.724&  2.1& Y& Y& 1& 3.550\\
spec-10228-58223-0625& 144.98375& 34.2290730& 3.598&  8.6& Y& Y& 1& 3.356\\
spec-10229-58441-0026& 146.63169& 34.8369130& 3.595&  5.8& Y& Y& 1& 3.398\\
\hline    \end{tabular}
    \caption{Summary of the properties of the first five SDSS/DR16 quasars included in the classification. We list: the quasar name in the SDSS convention, the right ascension, the declination, the quasar redshift from the SDSS pipeline; the $S/N$ measured at 1150~\AA\ in the quasar rest-frame; a flag indicating whether the quasar is a ``Science Primary" observation (Y or N); a flag indicating whether the quasar is included in the training set (Y or N); a flag describing the outcome of the classification (1: quasar with LLS; 2: quasar without LLS; 4: non quasar); the measured redshift of the LLS. The full table is available online.}
    \label{tab:llslist}
\end{table*}

\begin{figure}
    \centering
    \includegraphics[scale=0.55]{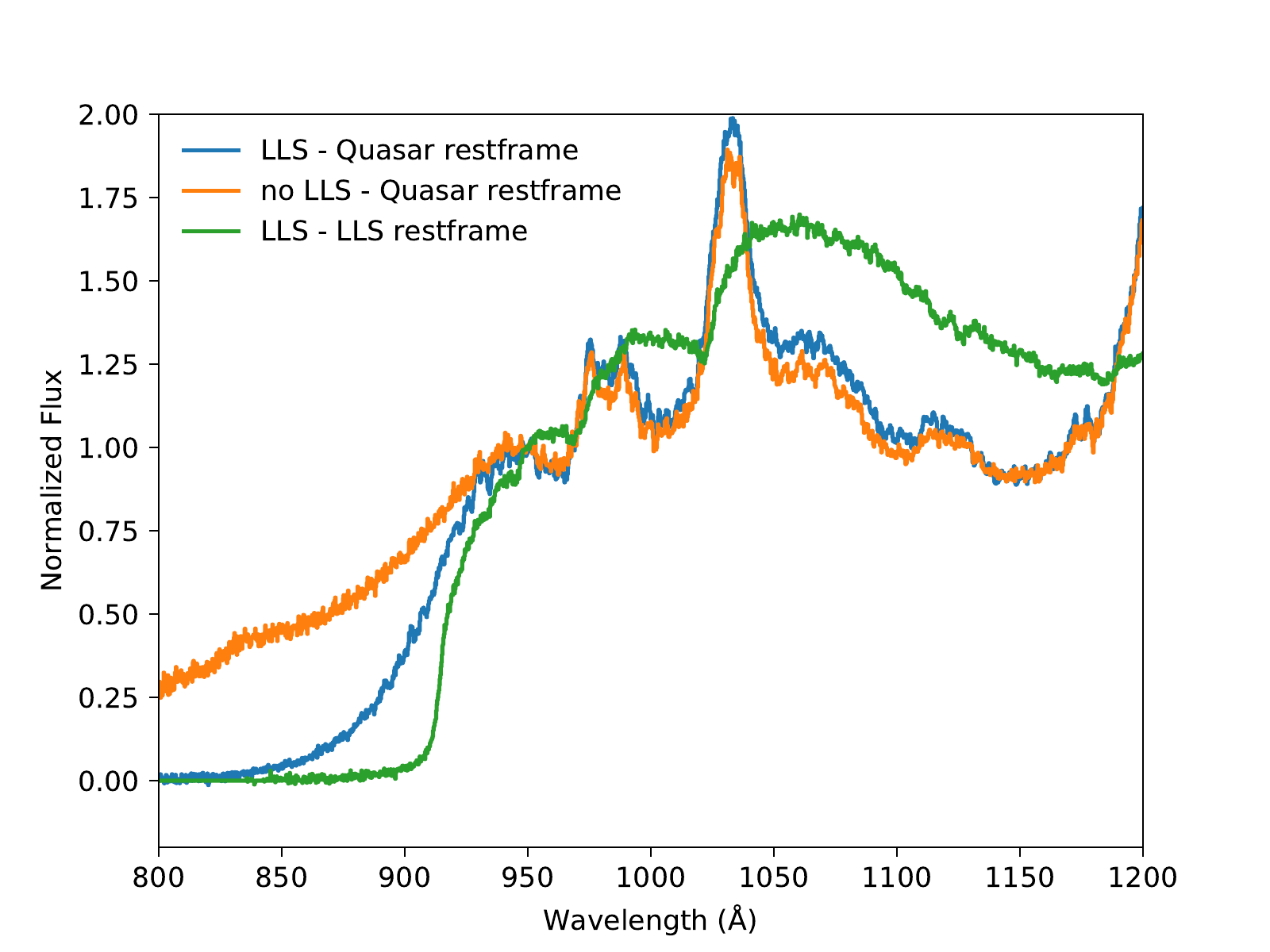}
    \caption{Median stack of quasar spectra, normalized at 950~\AA: i)
    containing LLSs in the absorber rest frame (green), ii) containing LLSs in the quasar rest frame (blue), iii) without LLSs in the quasar rest frame (orange). The random forest classifier produces  samples of quasars with and without LLSs with statistical properties that match expectations.}
    \label{fig:qsostack}
\end{figure}

\begin{figure*}
    \begin{tabular}{c|c}
       \includegraphics[scale=0.55]{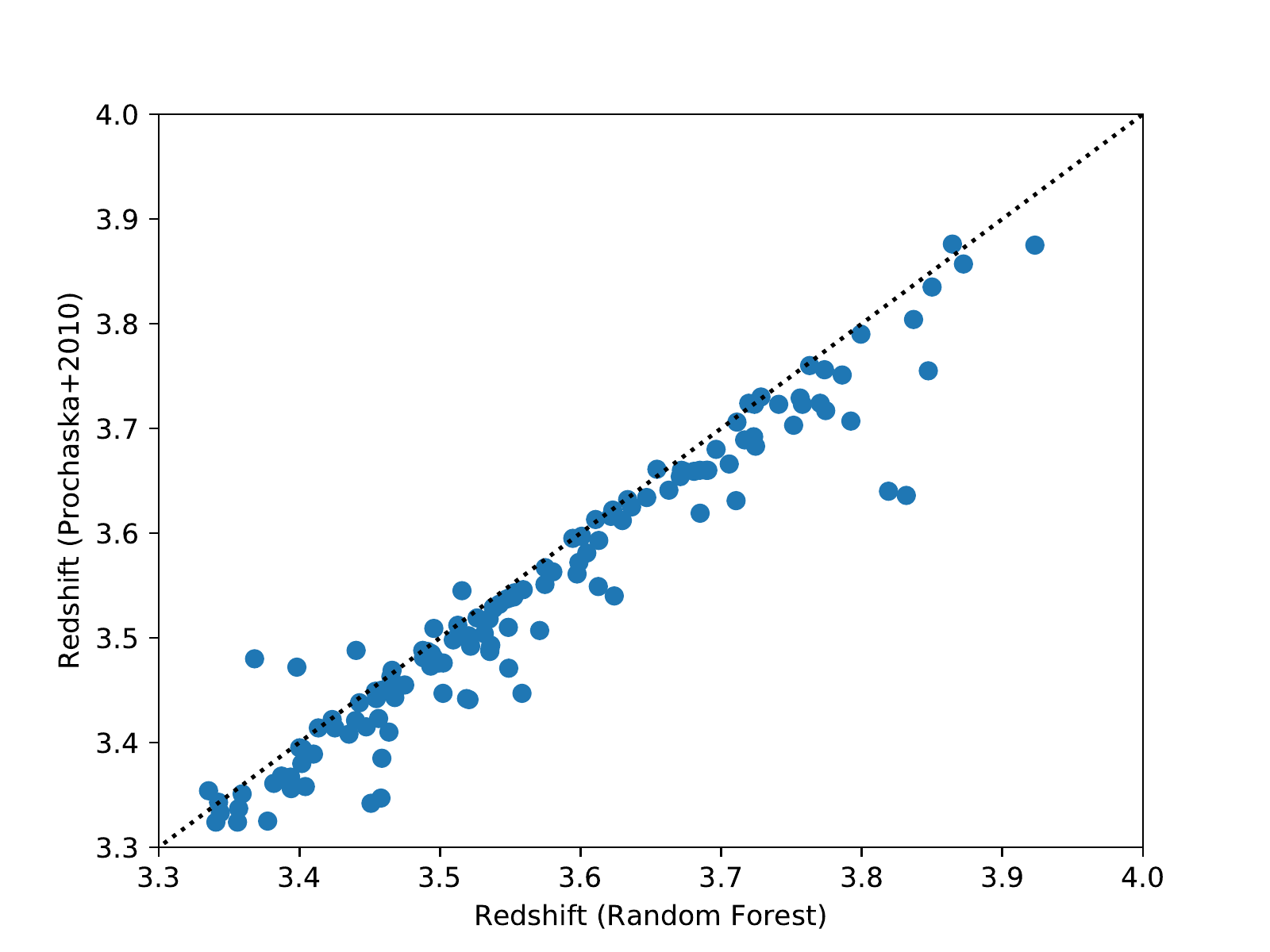}
     & \includegraphics[scale=0.55]{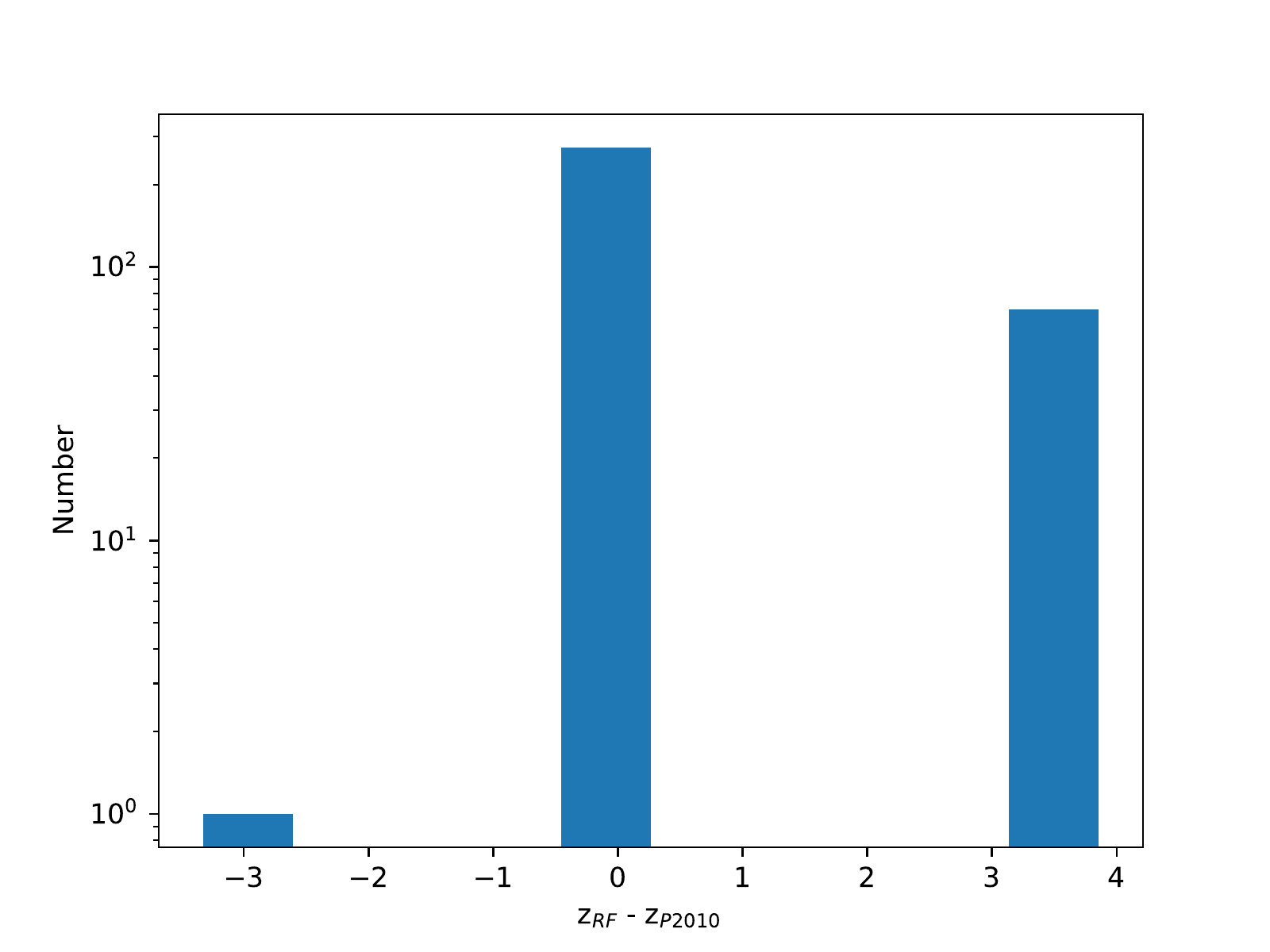}
    \end{tabular}
    \caption{Left: Comparison of the redshift determination for 273 LLSs at $z>3.3$ in common between our ML catalogue and the statistical sample by \citet{prochaska2010}. Right: Histogram of the discrepancy between the redshift determination of LLSs identified in 344 quasars in common between our ML catalogue and the statistical sample by \citet{prochaska2010}. By plotting the difference in redshift and assigning $z=0$ in case of no LLSs, sightlines for which agreement is found (both with and without LLSs) appear at $z_{\rm RF} - z_{\rm P2010}\approx 0$ (273/344 cases). Sightlines for which \citet{prochaska2010} identify a LLS which is not present in the ML catalogue appear at $z_{\rm RF} - z_{\rm P2010}\approx -3$ (1/344 cases). Conversely, LLSs in the ML catalogue that are not in the \citet{prochaska2010} statistical sample appear at  $z_{\rm RF} - z_{\rm P2010}\approx -3$ (70/344 cases).}
    \label{fig:cfrproch}
\end{figure*}

\begin{figure}
    \centering
    \includegraphics[scale=0.55]{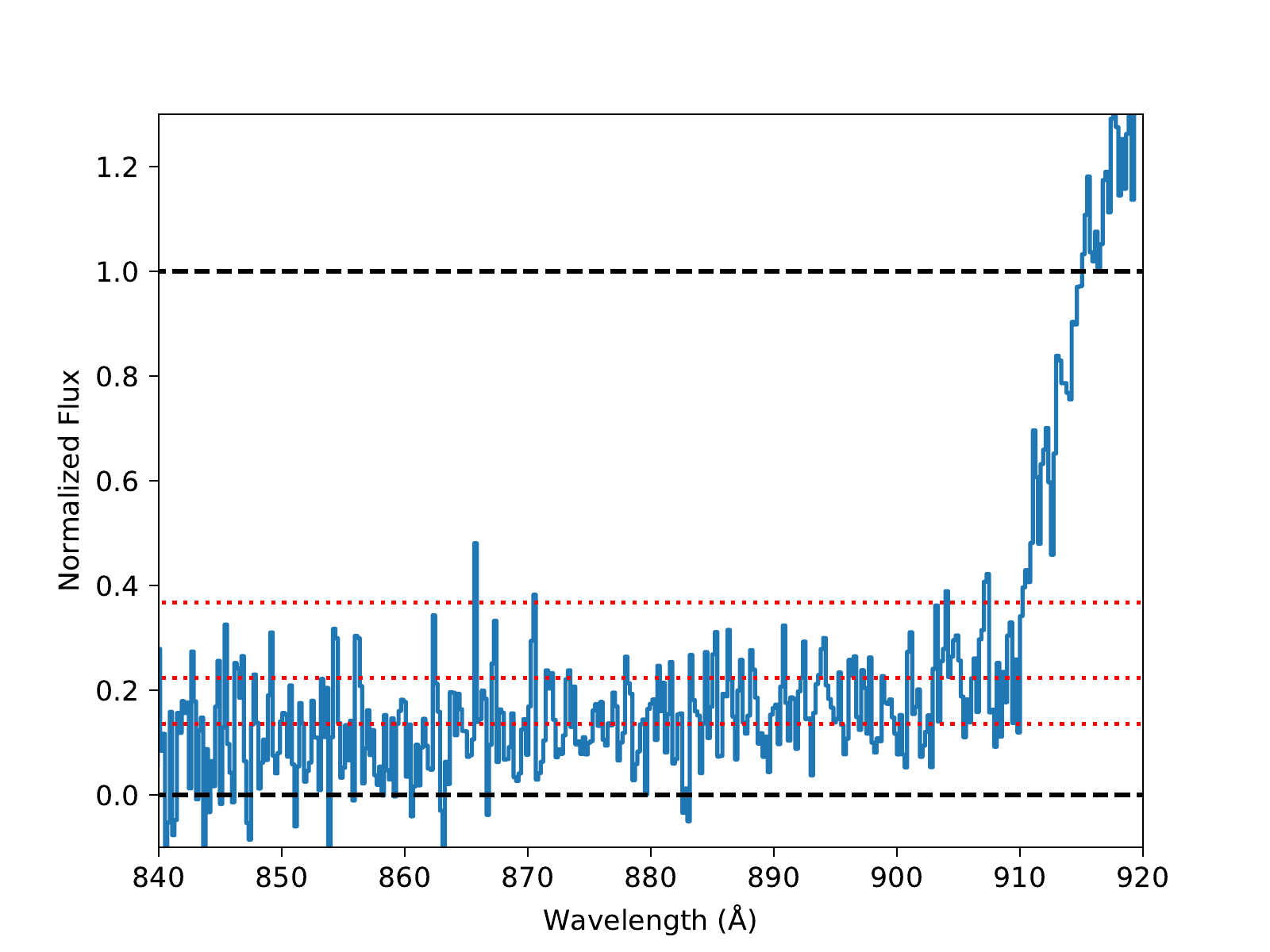}
    \caption{Median stack of 70 quasars in the LLS restframe for which we identify a LLS that is not present in the catalogue by \citet{prochaska2010}. Spectra are normalized in the range $913-918$~\AA. Dotted lines show, from top to bottom, the expected transmission for optical depths of 1, 1.5, and 2. Sightlines with a discrepant classification appear to host LLSs close to the $\tau \approx 2$ limit, and hence are more ambiguous to classify especially at low $S/N$.}
    \label{fig:stackdiff}
\end{figure}

As a last step, we vet the classification of the machine learning to verify that the above procedures yield a classification consistent with the expectation of a human classifier. We find that the training has been successful and LLSs are identified and redshifts are measured consistently to the expectation of a human classifier.  As is the case for the human classification, spectra at low $S/N$ are naturally subject to a more ambiguous classification. In principle, we could add a new class in the classifier, or even train the classifier to separate both ambiguous LLSs and ambiguous clear sightlines from confident classifications. However, due to the quite limited sample size of our training set, we defer this refined classification to future work based on larger samples of spectra, for example from WEAVE and DESI. For the time being, readers interested in a higher-purity sample should remove sightlines with $S/N<5.5$ per pixel between 1120~\AA\ and 1180~\AA\ in the quasar rest frame, which we find to be a good discriminant between confident and ambiguous classifications in our training set.  

More quantitatively, in Fig.~\ref{fig:qsostack} we show three median stacks of quasar spectra normalized at 950~\AA\ either in the quasar (blue and orange) or LLS (green) rest frame. The stack of sightlines with LLSs in the absorber's rest frame clearly shows a marked flux decrement at $\approx 912~$\AA, with a transmitted flux bluewards of the Lyman limit of $\approx 0$, as expected for $\tau \gtrsim 2$ LLSs. This confirms the general purity of the catalogue, in line with the scores obtained from the classifier. Also visible are the flux decrements associated with the IGM Lyman series, although no sharp absorption lines are visible. This effect is due to the smoothing introduced by the imprecise redshifts derived from the location of the Lyman limit (see e.g. Fig.~\ref{fig:llsprop} and Fig~\ref{fig:cfrproch}). A better redshift determination, e.g. based on the Lyman series lines in the training set, will improve the overall redshift accuracy.     

Examining instead the two stacks in the quasar rest frame (blue and orange lines in Fig.~\ref{fig:qsostack}), we note a substantial transmission below the quasar Lyman limit. A flux decrement is to be expected also when no LLSs are present, as systems with \HI\ column density below $10^{17.5}~\rm cm^{-2}$ account for $\approx 50\%$ of the opacity of the Universe at these wavelengths \citep[e.g.][]{prochaska2010,fumagalli2013}. Indeed, when restricting to sightlines with identified LLSs, a more marked flux drop is visible, although not as sharp as in the stack in the LLS rest frame, because of the smoothing introduced by the different LLS redshifts.

As a final validation of the catalogue, we compare our ML classification with the visual classification by \citet{prochaska2010}. By matching their ``statistical sample" (i.e. the optimal set of sightlines useful for the determination of the incidence of LLSs) with the ``Science Primary" sightlines in our catalogue, we identify 344 quasars in common. Among those, there are 273 LLSs in common at $z>3.3$, the redshift lower limit considered in \citet{prochaska2010}. 
Fig~\ref{fig:cfrproch} (left) shows a direct comparison of the redshift determination of these LLSs. A clear tight correlation is visible, possibly with a small bias towards higher redshift for the ML measurement with respect to the \citet{prochaska2010} one. Such a (small) discrepancy arises from the fact that the redshifts of the training sample have been obtained based purely on the position of the Lyman limit, which is intrinsically more imprecise than a redshift determination based on the Lyman series. Indeed, the scatter of this relation confirms a typical error of $\Delta z \approx 0.03$ for the ML redshift, assuming the \citet{prochaska2010} values as reference. The very few outliers arise from the ambiguity on the classification of partial LLSs in sightlines with multiple strong absorbers, especially at low $S/N$.

In the right hand-side of Fig~\ref{fig:cfrproch}, we assess more generally the consistency in the classification of the ML procedure and the visual classification by \citet{prochaska2010}. To this end, we measure the difference in redshift of LLSs in each sightline, assigning $z=0$ to sightlines with no LLSs. Sightlines for which there is agreement between the two classifications appear at $z_{\rm RF} - z_{\rm P2010}\approx 0$. Conversely,  at $z_{\rm RF} - z_{\rm P2010}\approx -3$ and $z_{\rm RF} - z_{\rm P2010}\approx 3$
 appear sightlines for which only one of the two methods identifies a LLS, with the sign differentiating the method that yields a positive identification. From this comparison, we find that for $273/344$ cases ($\approx 80\%$) the two methods agree on the classification. Only in $1/344$ instances, \citet{prochaska2010} identifies a LLS that is not included in the ML catalogue. Finally, in $70/344$ cases ($\approx 20\%$), we identify a LLS that is not included in the ``statistical" sample by \citet{prochaska2010}. To better understand the nature of this discrepancy, we study  in more detail the sightlines in this last class.

\begin{figure*}
    \centering
    \begin{tabular}{cc}
    \includegraphics[scale=0.55]{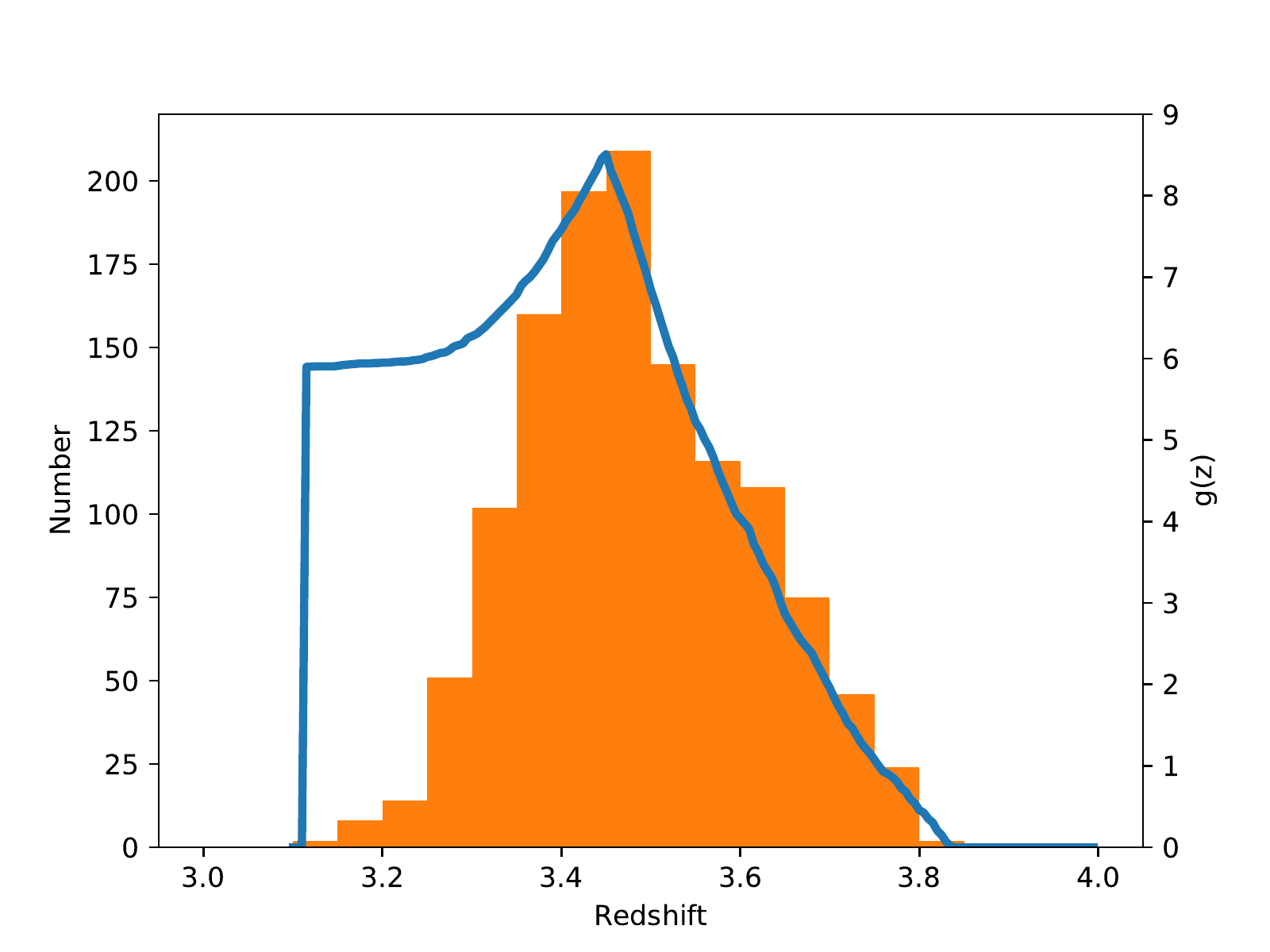}&
    \includegraphics[scale=0.55]{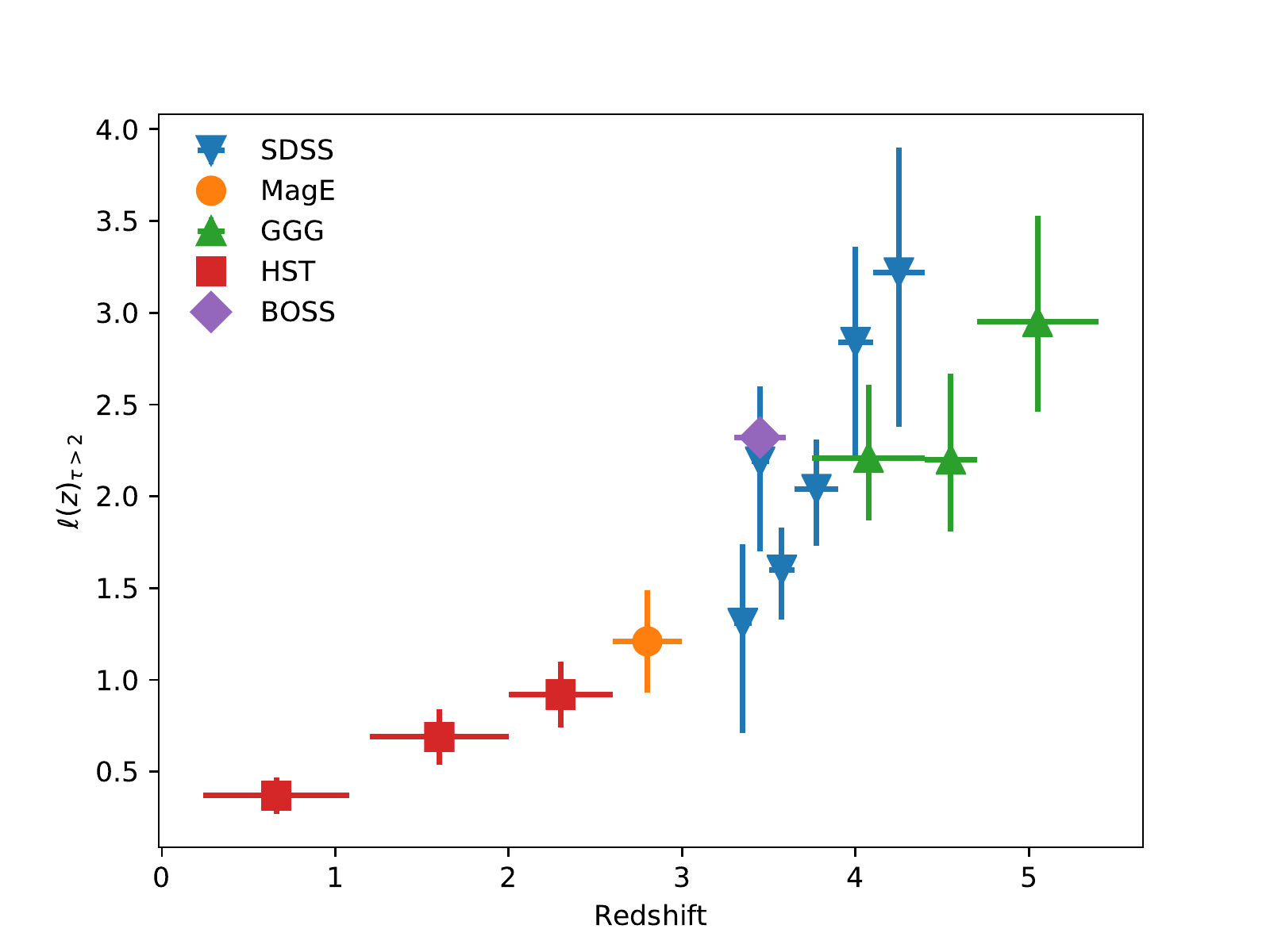}
    \end{tabular}
    \caption{Left: Histogram of the LLS redshifts for the ``Science Primary" quasars at $z\ge 3.6$ (orange). Also plotted, with values on the right hand-side, is the surveyed path length, $g(z)$. Right: The redshift evolution of the number of $\tau \ge 2$ LLSs from literature data (see text for references) and our ML based sample (purple diamond). Our classification yields a LLS sample characterized by a number per unit redshift that is in line with literature determinations.}\label{fig:lofz}
\end{figure*}

Fig.~\ref{fig:stackdiff} shows the median stack of these 70 quasars, in the rest frame of the LLSs identified with the ML procedure. Spectra are normalized redwards of the Lyman limit, in the interval of $913-918$~\AA. The subset of sightlines with discrepant classification appears in fact to host LLSs, although with residual transmission at the Lyman limit. While a precise measure of the optical depth of the stack spectrum is sensitive to the exact normalization, from  Fig.~\ref{fig:stackdiff} we establish that the stack is more absorbed than those hosting $\tau \approx 1$ LLSs, and is in fact similar to the ones with $\tau \approx 2$ LLSs. Thus, we conclude that the majority of the discrepant sightlines arise from LLSs that are close to the boundary of the classification at $\tau = 2$. These sightlines are more easily missed especially at the lower data quality of the original SDSS DR7 data used by \citet{prochaska2010}, so we argue that their catalogue may be somewhat incomplete close to $\tau = 2$. Similarly, an imprecise determination of the column density in our training set may include some  $\tau \lesssim 2$ LLSs, resulting in the inclusion of partial LLSs below $\tau = 2$ in the ML catalogue. Establishing which of the two effects dominate (and hence which catalogue is more complete/pure)  is non trivial, although it is likely that both effects are present at the same time, and the true answer is bracketed by the two catalogues.

\section{The physical properties of LLS\texorpdfstring{\MakeLowercase{s}}{s}}\label{sec:llsprop}

With a sample of new LLSs from SDSS/DR16, we revise the determination of the number of LLSs per unit redshift, $\ell(z) \equiv
\frac{dN}{dz}$. This metric, conceptually similar to the luminosity function for galaxies but specialized for measurements along the line of sight \citep[e.g.][]{bahcall1969}, offers a useful view of the redshift evolution of the population of LLSs, which can be linked to galaxy evolution models \citep[e.g.][]{fumagalli2013}.

For this calculation, we follow literature work \citep[e.g.][]{prochaska2010,fumagalli2013}. First, we restrict to the high-quality sample of ``Science Primary" sightlines with $S/N > 5.5$ at 1150~\AA. To avoid the colour bias identified in SDSS \citep[e.g.][]{prochaska2010,worseck2011}, we further restrict to quasars with redshift $z > 3.6$. We indeed verify that, without the inclusion of this redshift cut, $\ell(z)$ increases by $\approx 10\%$ at $z \approx 3.4$, and by $\gtrsim 20\%$ at $z\approx 3.3$.
We finally exclude sightlines with proximate LLSs, defined as those with redshift within 10,000~$\rm km~s^{-1}$ of the quasar redshift \citep[e.g.][]{perrotta2016}.

With this selection, we consider 2439 sightlines, containing 1259 LLSs. 
Next, we compute the useful pathlengh searched in all the sightlines, $g(z)$, defined as the sum of the redshift probed starting from 10,000~$\rm km~s^{-1}$ of the quasar redshift, up to the lowest redshift probed along each quasar. This lower bound is defined either by the bluest wavelength searched in our analysis ($3750~$\AA) or by the redshift of the first $\tau \ge 2$ LLS identified. Formally, this redshift is defined by $\max (3750/912-1,z_{\rm LLS})$.
Fig.~\ref{fig:lofz} (left) shows the probed path length, the shape of which is modulated by the quasar redshift distribution on the right hand-side of the $z\approx 3.4$ peak, and the redshift distribution of LLSs and the left of the peak. 

As evident from Fig.~\ref{fig:lofz}, most of the statistical power of our search is between $z=[3.3,3.6]$. Above $z>3.6$, the limited number of quasars result in a rapidly decreasing $g(z)$ that makes the final determination of $\ell(z)$ prone to very large completeness corrections. Below $z<3.3$, instead, the recovery of LLSs is more affected by the lower $S/N$ at the bluest wavelengths \citep[for a discussion on this effect, see also][]{prochaska2010}. 
Once restricted to the selected redshift interval, we compute $\ell(z)$ as the ratio of the number of LLSs in the range $z=[3.3,3.6]$ and the total redshift path $\Delta z = \int^{3.6}_{3.3} g(z) dz$, finding $\ell(z) = 2.32 \pm 0.08$. The error bar is based on Poisson statistics. 

In Fig.~\ref{fig:lofz} (right), we compare our new measurement with the results from the literature. Here, we show the low-redshift measurements based on {\it Hubble Space Telescope} data \citep[red square]{ribaudo2011,omeara2013}, the $z\approx 3$ measurement from the MagE survey by \citet[orange circle]{fumagalli2013}, the $z>3.3$ SDSS DR7 data from \citet[blue downward triangle]{prochaska2010}, and the high-redshift measurement from the Giant Gemini GMOS (GGG) survey by \citet[green triangle]{crighton2019}. 

Our determination agrees well with 
literature values, and it is curiously overlapping with the data point at comparable redshift by \citet{prochaska2010}. Given the small error bar due to the much larger sample size of our catalogue, however, our BOSS measurement is perhaps somewhat discrepant ($\approx 30\%$ higher) compared to the trend outlined by all the literature values as a function of redshift. 
As discussed in the previous section, this difference may be the result of two competing effects: first, the inclusion of LLSs just below  the $\tau = 2$ boundary in our catalogue and, second, the potential incompleteness at $\tau \approx 2$ LLSs in the catalogue by \citet{prochaska2010}. Modulo a better classification at the boundary of $\tau \approx 2$, future determinations in larger surveys like DESI and WEAVE have the potential to pin down reliably the redshift evolution of LLSs above $z>3$. 

\begin{figure}
    \centering
    \includegraphics[scale=0.55]{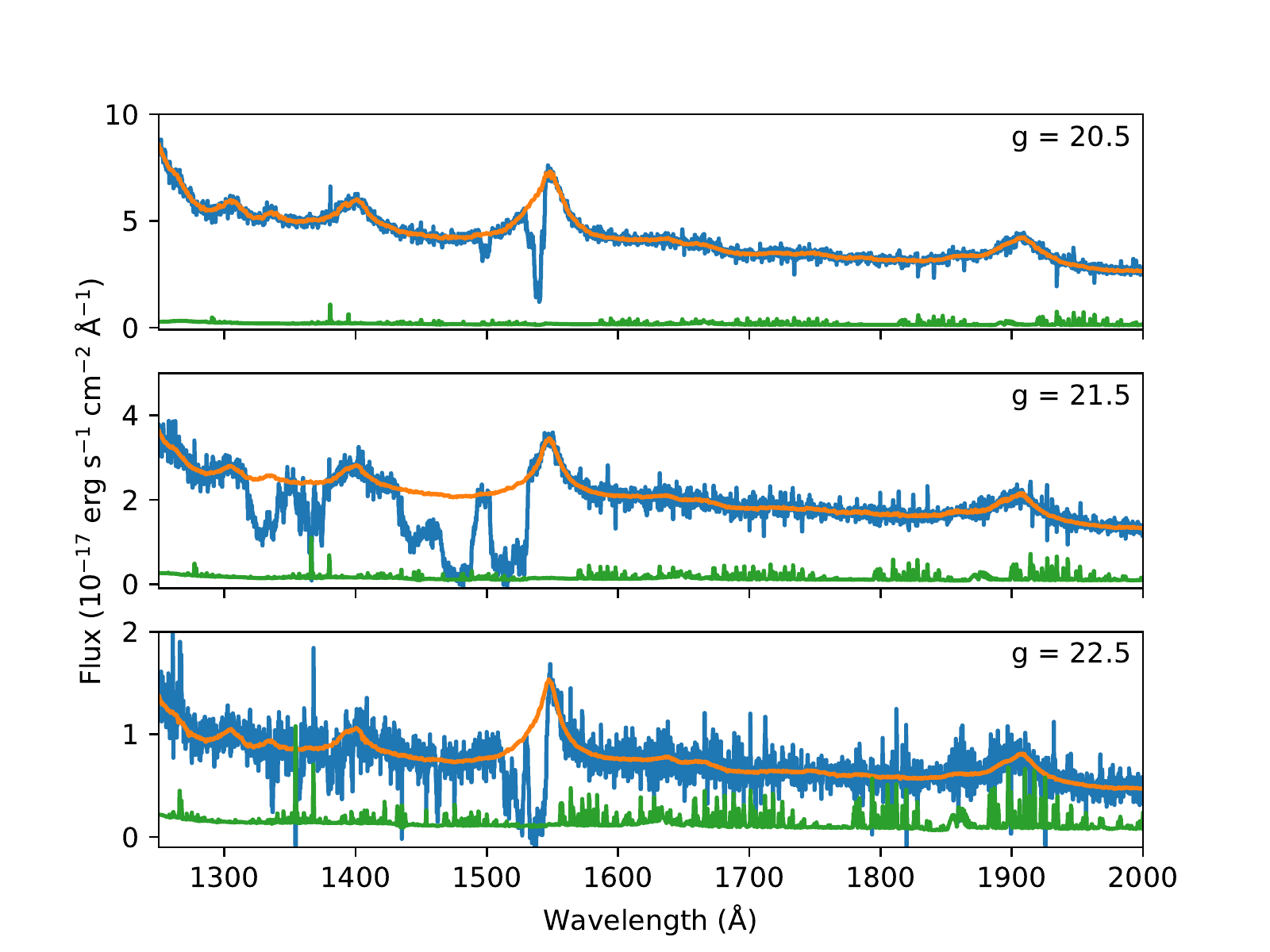}
    \caption{Examples of continuum modelling for three random mock spectra including BAL features ($t_{\rm exp} = 4000~$s). The flux is shown in blue, the flux error is shown in green and the model continuum is shown in orange. Our procedure (see text for details) performs well in the region of interest ($1300-1600$~\AA) even in the presence of deep BAL features.}
    \label{fig:balcontinuum}
\end{figure}

\begin{figure}
    \centering
    \includegraphics[scale=0.55]{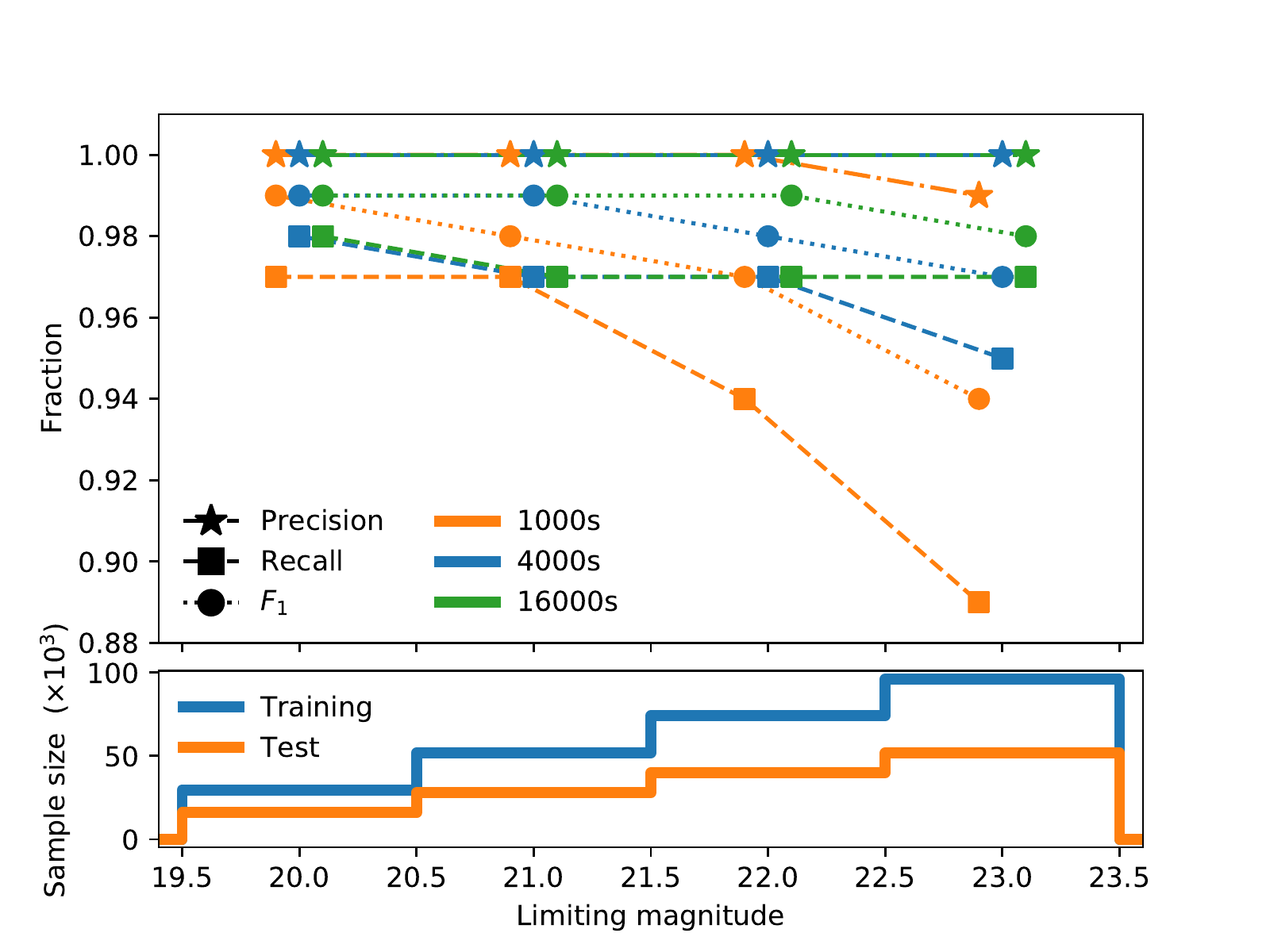}
    \caption{Same as Fig.~\ref{fig:llsstats}, but for the classification of BALs in mock data. At sufficiently high
    $S/N$ (achieved for $t_{\exp} \ge 4000$~s), the recovery of BALs has high completeness and purity independently of the quasar magnitude.}
    \label{fig:balstats}
\end{figure}

\section{Classification of BAL quasars}\label{sec:mockbal}

Studies of intervening absorption line systems often rely on the exclusion of BAL quasars from the sample, as the BAL features  affect a substantial part of the path-length and could confound some of the signatures (e.g. damped profile) of intervening absorbers.
While dedicated efforts to identify BAL quasars with machine learning can be found in the literature  \citep[e.g.][]{busca2018,guo2019}, we conclude this work by exploring the performance of the off-the-shelf implementation of random forest
in the identification of BALs. 

 For this task, we rely again on the implementation of the random forest classifier in \scikit, and test the effectiveness of the classification using mocks that include BALs. 
 The procedure we follow is similar in spirit to the one developed for the search of LLSs. 
 First, quasar spectra are aligned in the rest-frame through a linear interpolation. To reduce the complexity of the problem (i.e. we wish to classify the presence of BAL feature without the need to learn the intrinsic shape of the quasar spectrum), we continuum normalize the spectra using a non-negative matrix factorization (NMF) as in \citet{zhu2013}. We note that, for mocks, this procedure is to some extent redundant, as the quasars themselves have been constructed from a set of eigenvectors derived via principal component analysis which are then combined in a linear combination \citep[e.g.][]{paris2012}.  However, by recomputing the continuum on the mock spectra we approximate more closely the actual analysis that can be performed on real data.  
 
To continuum normalize the quasars, we first decompose a set of 25,000 mock quasars with $m_{\rm g} < 21~\rm mag$ without BALs into 12 positive components, where the number of components is chosen in line with previous work \citep{paris2012,zhu2013,guo2019}. We then fit a positive linear combination of these components to the entire sample of 180,000 mock quasars, also including objects with BALs. To avoid the need for masking BAL features, which can affect the quality of the resulting continuum, we only fit each spectrum redwards of the \ion{C}{IV} line ($\lambda>1550~$\AA) and extrapolate the best-fit model in the region bluewards to \ion{C}{IV}. As shown in Fig.~\ref{fig:balcontinuum}, we are able to accurately predict the continuum level of quasars with BALs down to faint magnitude limits.   

Once spectra are continuum normalized, we construct a $n_{\rm qso} \times n_{\rm pix}$ flux image, where the number of pixels is chosen to cover the wavelength range $\approx 1396-1600$~\AA. This window brackets the \ion{C}{IV} region of the spectrum that contains the \ion{C}{IV} BALs at $\lambda=1548$~\AA\ and $\lambda=1550$~\AA, which are the features we aim to classify. We prefer to focus on the \ion{C}{IV} region, as this is a cleaner region of the spectrum compared to hydrogen lines embedded in the \lya\ forest. As fluxes are already in the range (0,1), we do not apply further regularization of the data. 

We note that quasars may contain additional features in this region of the spectrum, most notably metal lines associated with strong absorption line systems such as DLAs. As our mocks are approximated and do not contain detailed modelling of the metal lines arising from associated and intervening absorbers, the precision of our classification is expected to be optimistic, as for instance strong \ion{C}{IV} lines from proximate DLAs may be confused for BAL features. 
We foresee, however, that the majority of spectra will be still correctly classified, as DLAs only rarely exhibit very complex kinematics in excess of $\approx 500~\rm km~s^{-1}$ as is typical instead of many BALs \citep{fox2007,trump2006}.

Following the preparation of the data, we move to the classification problem using the {\sc RandomForestClassifier} method in \scikit. First, we perform a grid search to tune the parameter of the classifier using a five-fold cross-validation with $F_1$ as score method on a training set containing 65\% of quasars with $m_{\rm g} < 21.5~$mag. We find that the best performance is obtained for {\tt n\_estimators=500}, {\tt criterion=`gini'}, {\tt min\_samples\_split=2}, and {\tt max\_features=50}. However, as for the case of LLSs, we find that the classifier is only weakly sensitive to the exact choice of parameters. 

Next, we apply the classifier to all mocks. To study the performance as a function of magnitude, we iterate the procedure for $g-$band magnitude limits of 20, 21, 22, and 23~mag. At each iteration, we train the classifier on a set containing 65\% of the total sample size, and test the performance on the remaining objects. A summary of the sample size, precision, recall, and $F_1$ score for each case is presented in Fig.~\ref{fig:balstats}.
As seen from Fig.~\ref{fig:balstats}, the classification is very precise in recognizing BALs, with little to no confusion with other features. This is insensitive with respect to limiting magnitude and, largely, $S/N$ of the data that in this region of the spectrum is generally high ($S/N \gtrsim 3$ also for the fainter quasars). As described above, the actual precision is expected to be lower in real data which contain additional absorption lines in this region of the spectrum.

The recall of true BALs, while not as high as the precision, is consistently around $\approx 97\%$, with a more marked dependence on $S/N$. This is particularly noticeable for the case of $t_{\rm exp} = 1000~$s,
where the fraction of true BALs identified quickly decreases with magnitude. Looking at the false negative, the classifier tends to miss primarily BALs which are associated with the weakest \ion{C}{IV} features with rest-frame equivalent widths $<5~$\AA, implying high completeness with respect to typical BALs with equivalent widths $\gtrsim 10~$\AA.
We therefore conclude that the random forest classifier can provide a valid alternative to the convolution neural network classifiers that have been developed in the literature.

\section{Summary and Conclusions}\label{sec:concl}

We present the development of a pipeline designed to identify and measure the physical properties of $\tau \ge 2$ LLSs in large surveys. Differently from previous searches for strong absorption line systems like DLAs that target the characteristic shape of the Ly$\alpha$ absorption line, we rely on the distinctive flux drop at the Lyman limit around 912~\AA\ in the system's rest frame (see Fig.~\ref{fig:llsfeatures}). Algorithmically, we choose the random forest implementation in \scikit\, using the classifier for the identification of LLSs and the regressor for the measurements of physical properties (redshift and column density).
To test the reliability of the results and to develop the pipeline, we build a library of 180,000 mock quasar spectra including DLAs and LLSs, with data quality comparable to what achievable in upcoming surveys such as DESI and WEAVE.  

After aligning quasar spectra and performing continuum normalization in the region bluewards of the quasars' Lyman limit, we optimize the parameters that control the random forest classifier using a  five-fold  cross-validation  with $F_1$ score as metric  of  performance, finding that the method is rather insensitive to the choice of parameters with  $F_1\approx 0.9$ for all reasonable variations.  After training the classifier on $65\%$ of the mock dataset, we apply the classification on the remaining mock data. The classification is sensitive to the spectral $S/N$, with precision and recall increasing as a function of decreasing quasar magnitude and increasing exposure time (Fig.~\ref{fig:llsstats}). Once binned to 20 times the original sampling, we find recall and precision above $90\%$ for typical DESI or WEAVE like spectra with $g\lesssim 23~$mag.  

We then proceed to test the ability of the random forest regressor in measuring the redshift and \HI\ column density of the identified LLSs. For this step, we follow a data pre-processing that is similar to the one of the previous step. We find that the regression is able to recover accurately the LLS redshift, with discrepancies from the true value generally within $\Delta z \lesssim 0.025$. The recovery is again sensitive to the spectral $S/N$, with an $R_2$ coefficient that depends on quasar magnitude and exposure time and that is above 0.8 for DESI and WEAVE like spectra (Fig.~\ref{fig:llszfit}). 
Conversely, we find that the random forest regression performs poorly for the recovery of column density. This is not unexpected, as the information of the column density is encoded in the depth of the flux decrement at the Lyman limit, a quantity that quickly saturates for $\tau > 2$ systems. Thus, simply by examining the flux at the Lyman limit, there is insufficient information to  reliably pin down the column densities. 

We next proceed to apply this pipeline on 10,000 real quasar spectra between $z\approx 3.5-4.0$ from SDSS/DR16. After pre-processing the data as done for the mocks and cleaning the catalogue for a small fraction of spurious (non quasar) sources, we train the random forest classifier on a sample of $\approx 2,000$ spectra that we visually inspect and classify to identify the presence of LLSs at $z>3.11$. In line with the performance obtained with the mock observations, we find an $F_1\approx 0.88$ and a recall and precision of 0.92 and 0.94 for LLSs, respectively, on real data. Applied to the entire sample, we identify $\approx 6600$ LLSs, a number that reduces to $\approx 4800$ once we remove duplicate observations. 
Using a random forest regression, we further measure the redshift of the LLSs. 

With this catalogue in hand, we perform several tests to validate the pipeline, including the analysis of stacks of sightlines hosting LLSs or not (Fig.~\ref{fig:qsostack}) and a detailed comparison with previous catalogues based on visual classification (Fig.~\ref{fig:cfrproch}). The method we have developed appears to perform well, yielding a catalogue that is consistent with previous classification and that has statistical properties in line with known properties of LLSs.
With this new catalogue, we finally assess the number of LLSs per unit redshift in the interval $z=[3.3,3.6]$, finding $\ell(z)=2.32\pm 0.08$ in line with previous literature determinations (Fig.~\ref{fig:lofz}). 

As a final exercise, we study the efficiency of the random forest classifier in the identification of BAL quasars. For this, we rely on a mock catalogue of quasars in which we inject BAL absorption lines following a set of pre-compiled templates. After removing the quasar continuum with a NMF decomposition (Fig.~\ref{fig:balcontinuum}), we tune the classifier with a five-fold cross-validation, finding again a very weak dependence of the performance on the parameter choice. Following the training on a training set, we study the performance of the classifier on a new set of mocks, finding very high precision and recall ($\gtrsim 97\%$, except for the lowest $S/N$ quasars). 

In conclusion, building on a off-the-shelf implementation of a random forest classifier and regressor in \scikit, we have developed an efficient pipeline for the recovery of LLSs and the measurement of their redshifts, purely based on the characteristic flux decrement at the Lyman limit. Following the successful validation of this technique using both mock and real data in comparison with previous work, we have obtained a framework that can now be applied to upcoming large surveys like DESI and WEAVE. We are therefore well positioned to produce catalogues of tens of thousands of LLSs, with which we can study in detail the redshift evolution and other properties such as the metallicity of strong absorption line systems, which are the ideal systems to study the galaxy CGM and the denser IGM in the high-redshift Universe. 

\section*{Acknowledgements}
Our thanks to Andreu Font Ribera, David Kirkby, Paul Martini, and John Moustakas for the development of the mock-making codes used in the paper.   
 SF acknowledges the financial support of the Swiss National Science Foundation. This project has received funding from the European Research Council (ERC) under the European Union's Horizon 2020 research and innovation programme (grant agreement No 757535). 
This work has been supported by Fondazione Cariplo, grant number 2018-2329.
This work used the DiRAC@Durham facility managed by the Institute for Computational Cosmology on behalf of the STFC DiRAC HPC Facility (www.dirac.ac.uk). The equipment was funded by BEIS capital funding via STFC capital grants ST/K00042X/1, ST/P002293/1, ST/R002371/1 and ST/S002502/1, Durham University and STFC operations grant ST/R000832/1. DiRAC is part of the National e-Infrastructure. 
We made use of Sloan Digital Sky Survey IV. Funding for the Sloan Digital Sky Survey IV has been provided by the Alfred P. Sloan Foundation, the U.S. Department of Energy Office of Science, and the Participating Institutions. SDSS-IV acknowledges
support and resources from the Center for High-Performance Computing at
the University of Utah. The SDSS website is \url{www.sdss.org}.
SDSS-IV is managed by the Astrophysical Research Consortium for the 
Participating Institutions of the SDSS Collaboration.

\section*{Data availability}
The classification of the SDSS/DR16 spectra is available in the online supplementary material. For access to other data products including mock spectra and codes used in this work, please contact the authors or visit \url{http://www.michelefumagalli.com/codes.html}.










\bsp	
\label{lastpage}
\end{document}